\documentclass[12pt]{iopart}

\expandafter\let\csname equation*\endcsname\relax
\expandafter\let\csname endequation*\endcsname\relax

\usepackage{caption}
\usepackage{subcaption}
\usepackage{graphicx}               
\usepackage[utf8]{inputenc}         
\usepackage{xstring}
\usepackage{soul,todonotes}
\usepackage[noabbrev]{cleveref}
\crefformat{equation}{equation~#2#1#3}

\newcommand{\kindex}[2]{\ensuremath{{#1}_{\scalebox{0.5}{#2}}}}
\usepackage[locale=UK,
number-mode=math,
unit-mode=text,
per-mode=symbol,
number-unit-product=\ ,
retain-zero-exponent=true]{siunitx}
\DeclareSIUnit{\pixel}{px}
\DeclareSIUnit{\fps}{fps}

\usepackage{etoolbox} 
\robustify{\kindex}

\graphicspath{{./figs/}{../figurematter/latex/figs/}}

\begin{document}
	\title[Phenomenology and scaling of optimal flapping wing kinematics]{Phenomenology and scaling of optimal flapping wing kinematics}
	\author{Alexander Gehrke and Karen Mulleners}
	\address{École polytechnique fédérale de Lausanne, Institute of Mechanical Engineering,
		Unsteady flow diagnostics laboratory, 1015 Lausanne, Switzerland}
	\ead{karen.mulleners@epfl.ch}
	\vspace{10pt}

	\begin{abstract}
		Biological flapping wings fliers operate efficiently and robustly in a wide range of flight conditions and are a great source of inspiration to engineers.
		The unsteady aerodynamics of flapping wing flight are dominated by large-scale vortical structures that augment the aerodynamic performance but are sensitive to minor changes in the wing actuation.
		We experimentally optimise the pitch angle kinematics of a flapping wing system in hover to maximise the stroke average lift and hovering efficiency with the help of an evolutionary algorithm and in-situ force and torque measurements at the wing root.
		Additional flow field measurements are conducted to link the vortical flow structures to the aerodynamic performance for the Pareto-optimal kinematics.
		The optimised pitch angle profiles yielding maximum stroke-average lift coefficients have trapezoidal shapes and high average angles of attack. These kinematics create strong leading-edge vortices early in the cycle which enhance the force production on the wing.
		The most efficient pitch angle kinematics resemble sinusoidal evolutions and have lower average angles of attack.
		The leading-edge vortex grows slower and stays close-bound to the wing throughout the majority of the stroke-cycle.
		This requires less aerodynamic power and increases the hovering efficiency by \SI{93}{\percent} but sacrifices \SI{43}{\percent} of the maximum lift in the process.
		In all cases, a leading-edge vortex is fed by vorticity through the leading edge shear layer which makes the shear layer velocity a good indicator for the growth of the vortex and its impact on the aerodynamic forces.
		We estimate the shear layer velocity at the leading edge solely from the input kinematics and use it to scale the average and the time-resolved evolution of the circulation and the aerodynamic forces.
		The experimental data agree well with the shear layer velocity prediction, making it a promising metric to quantify and predict the aerodynamic performance of the flapping wing hovering motion.
	\end{abstract}
	\submitto{\BB}

	%
	%
	%
	%
	%

	\section{Introduction}%
	Bio-inspired mechanical flapping wing systems have been increasingly used in the past decades to study and understand the behaviour of natural fliers and serve as inspiration for the design of flapping wing micro air vehicles (MAV)~\cite{ellington_leading-edge_1996, sane_control_2001, lehmann_aerodynamic_2005, chin_flapping_2016}.
	Recently, MAV with similar sizes and weights as natural fliers have found their applications~\cite{keennon_development_2012, phan_design_2017, karasek_tailless_2018, jafferis_untethered_2019}.
	With the development of novel wing actuators~\cite{haider_recent_2020} and the miniaturisation of flight control systems and improvements in energy storage, MAV are employed to accomplish complex autonomous missions in urban environments~\cite{elbanhawi_enabling_2017}.
	With the decrease in size, the Reynolds number reduces and unsteady effects have more influence on the aerodynamic performance of the fliers.
	At lower Reynolds numbers ($Re < 5000$), flapping wing vehicles generally perform better than revolving wing aircraft and at $Re < 100$ the lift-to-power ratio is about twice as high for flapping wings in comparison to their revolving counterparts~\cite{pesavento_flapping_2009, zheng_comparative_2013}.

	Natural flapping wing fliers are extremely versatile.
	They seamlessly change between hovering and forward flight, use their wings to generate both lift and thrust, and can even glide to conserve energy.
	Flapping wings operate at high angles of attack above the static stall angle of the wing.
	These high angles cause a shear layer to separate at the leading edge which rolls up and forms a large scale coherent structure, the leading edge vortex.
	The stall of the wing is delayed through the rotational acceleration of the flapping wing which stabilises the leading edge vortex during the majority of the stroke cycle~\cite{lentink_rotational_2009}.
	A bound leading edge vortex creates a low pressure region on the suction side of the wing which generates high aerodynamic forces and torques required for the fast maneuverings of flapping wing fliers~\cite{ellington_aerodynamics_1984b, dickinson_unsteady_1993}.
	The unsteady aerodynamic effects of the leading edge vortex give rise to exceptional lift and thrust yields well beyond the aerodynamic performance of fixed wings under steady-state conditions~\cite{sun_high-lift_2005, jones_unsteady_2010}.
	At the end of the flapping half-cycle, the wing rotates to keep the leading edge in front of the trailing edge along the stroke direction.
	During the end-of-stroke rotation, the vortex separates from the shear layer and sheds into the wake and a new stroke begins.

	Nature's flapping wing fliers do not cease to amaze us with their incredible flight performance and efficiency, but many bio-inspired human-engineered devices do not yet manage to compete with their natural inspirers~\cite{mayo_comparison_2010}.
	One reason for this is that the functional morphology of insect wings is not yet fully understood and can not directly be incorporated in robotic flapping wing vehicles.
	During the natural evolution of birds and insects, the wing shape and their kinematics advanced simultaneously and different wing shapes favour specific kinematics for hovering flight~\cite{chin_flapping_2016}.
	Complex flapping wing motions are observed in nature~\cite{fry_aerodynamics_2005, bomphrey_smart_2017} and especially the pitch angle profile is highly depended on the wing geometry and elasticity but also varies with the flight conditions or flow characteristics expressed by the Reynolds number and reduced frequency~\cite{liu_size_2009, Matthews.2018}.
	Recent improvements in miniature wing actuators motivate the exploration of the influence of more complex wing kinematics on the flapping wing performance~\cite{haider_recent_2020}.

	Wing kinematics measured on birds and insects provide a starting point to design effective flapping wing motions but they are specific to each wing's properties and actuation system.
	Various parameter studies have been carried out in the past to characterise the performance of flapping wing kinematics for different wing planforms~\cite{wang_vortex_2000, sane_control_2001, phillips_effect_2011, bhat_effects_2020}.
	On a dynamically scaled mechanical model of a fruit fly Sane and Dickinson~\cite{sane_control_2001} varied the stroke amplitude, angle of attack, flip timing, flip duration and the shape and magnitude of stroke deviation in an extensive parameter study.
	Among other findings, they concluded that the mean drag increases monotonically with increasing angle of attack and a short flip duration advanced of the stroke reversal is beneficial for lift production.
	The influence of different stroke- and pitch angle waveforms at a fixed flapping frequency and amplitude was investigated recently by Bhat et al.~\cite{bhat_effects_2020} for a fruit fly wing planform.
	The stroke angle evolution was modulated between a sinusoidal and a triangular profile and the pitch angle evolution between a sinusoidal and a trapezoidal profile.
	The stroke angle evolution has a main influence on the magnitude of the lift coefficient \kindex{C}{L} maxima whereas the pitch angle evolution mostly impacts the instantaneous \kindex{C}{L} at stroke reversal.

	The vast parameter space of possible complicated flapping wing kinematics makes it challenging to derive general relationships between motion parameters and optimal aerodynamic performance.
	Experimental and numerical optimisations can aid to find optimal kinematics within the vast parameter space of the flapping wing actuation.
	Optimisations have been applied primarily to numerical models which are only limited by the computational cost and the validation of the numerical method~\cite{berman_energy-minimizing_2007, taha_wing_2013, lee_optimization_2018}.
	A hybrid optimisation approach which combines aspects of a genetic algorithm and a gradient-based optimiser was applied by Berman and Wang~\cite{berman_energy-minimizing_2007}.
	They parameterised the stroke, pitch, and elevation angle profiles to minimise the power usage on three differently weighted insect models in hovering flight.
	The aerodynamic forces are computed using a quasi-steady model and assuming a thin flat plate wing.
	The optimal kinematics found in their study exhibit a sinusoidal stroke evolution where the pitch angle is kept constant throughout the cycle.
	The kinematic functions found take advantage of passive wing rotation by using the aerodynamic moments to reverse the wing pitch.
	By treating the flapping wing kinematics optimisation as a calculus-of-variation problem along with quasi-steady aerodynamics, Taha et al.~\cite{taha_wing_2013} find that a triangular waveform for the stroke angle and a constant pitching angle throughout the half-stroke yield the best performance index in terms of $\kindex{C}{D}^2/\kindex{C}{L}^3$ with \kindex{C}{D} the drag coefficient.
	A stroke profile with a harmonic waveform requires \SI{20}{\percent} more aerodynamic power compared to the triangular waveform for the same performance evaluation.
	More recently, Lee and Lua~\cite{lee_optimization_2018} used a two-stage optimisation algorithm to investigate the effects of more complex, insect-like pitch angle kinematics on the hovering flight of a hawkmoth.
	They initiate the optimisation with a semi-empirical quasi-steady model to narrow down the parameter space and then use a computational fluid dynamics simplex optimisation method to refine the optimal pitch angle kinematics found.

	Quasi-steady or low-order unsteady aerodynamic models have good computational performance, however they are often restricted to wing kinematics within their local validated trajectory space.
	Computational fluid dynamics simulations at low $Re$ can accurately calculate the aerodynamic loads generated by a flapping wing, but are too computationally expensive to use in large scale optimisations.

	Experimental optimisations with dynamically scaled wings and force measurements combine accurate measurements with comparatively low experimental times~\cite{milano_uncovering_2005, martin_experimental_2018, gehrke_genetic_2018}.
	Automated data transfer and processing between the experimental system and the optimisation framework is required and the mechanism needs to have a robust control scheme and mechanical design to conduct a large number of iterations without human supervision.
	The early work of Milano and Gharib~\cite{milano_uncovering_2005} on experimental flapping wing optimisations applies a genetic algorithm to a two-axis system of a translating and rotating wing.
	The solution that yields the most lift in hovering flight was related to the strongest leading edge vortex growth.
	Martin and Gharib~\cite{martin_experimental_2018} employed a covariance matrix adaptive evolutionary strategy to find effective kinematics for a bio-inspired flapping fin which can be used as a side or a rear propulsor for underwater vehicles.

	In this study, we propose a unique robust optimisation scheme to obtain optimal pitch angle kinematics for a given wing geometry and Reynolds number on an experimental flapping wing platform.
	We employ a multi-objective evolutionary algorithm to find complex flapping wing motions which yield highest stroke-average lift and highest efficiency during hovering.
	The trade-off between lift and efficiency of the optimal solutions is represented by a Pareto front.
	Complementary velocity flow field measurements are conducted for the Pareto optimal kinematics to determine the leading edge vortex circulation and its position throughout the flapping wing cycle.
	The results consist of two parts.
	In a first part, we focus on explaining the interaction between the complex motion kinematics and the resulting aerodynamic performance using flow field data and qualitative information on the state of the leading edge vortex development.
	In the second part, we quantitatively describe and propose a novel approach to scale the temporal evolution of the vortex development and the aerodynamic forces and efficiency for all solutions along the Pareto front.
	%
	\section{Materials and Methods}
	\subsection{Wing Model and Kinematics}
\begin{figure}
		\centering
		\includegraphics[]{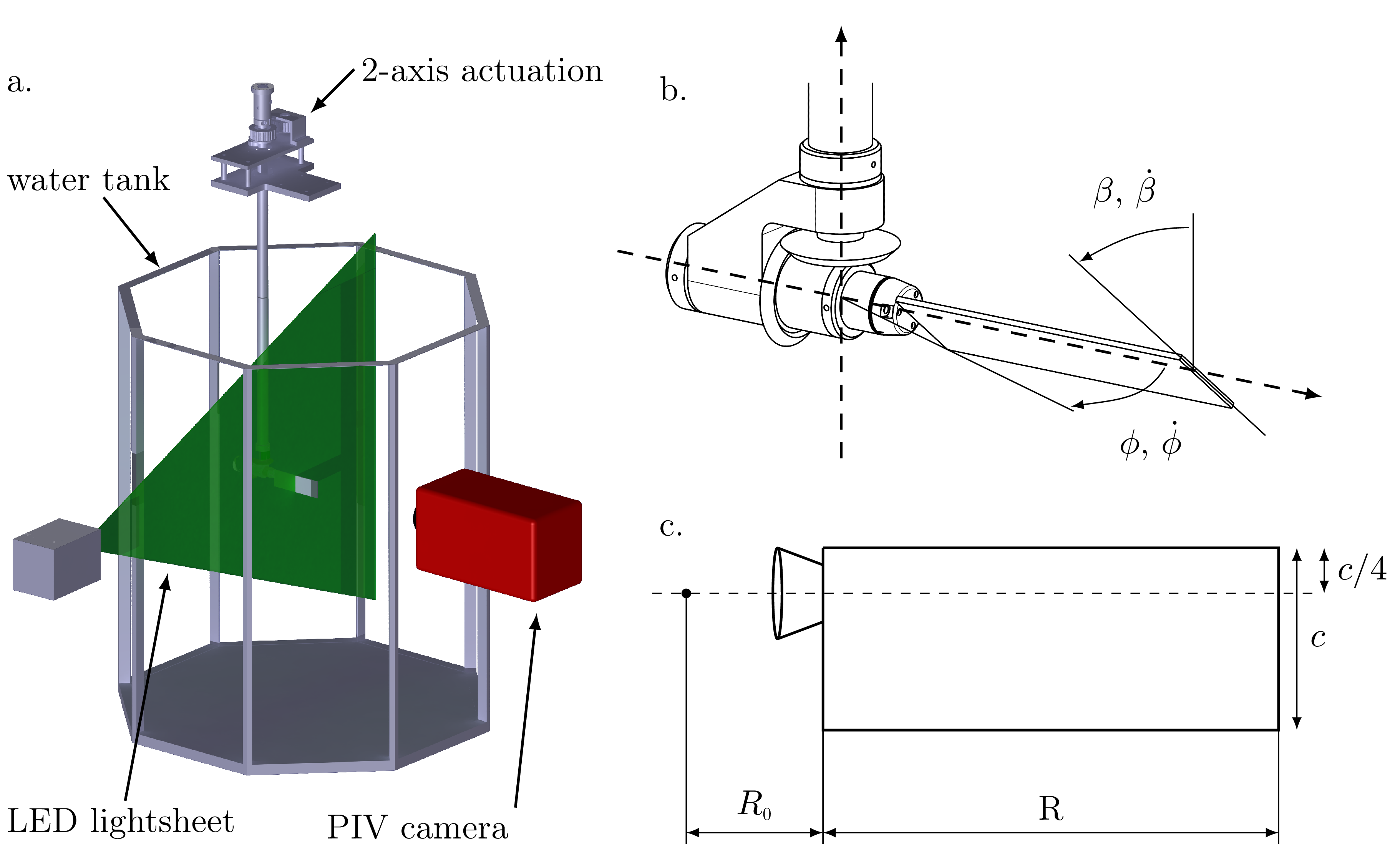}
		\caption{a. Schematic of the experimental configuration with the flapping wing mechanism submerged in an octagonal water tank.
			A light-sheet and camera are positioned to record the velocity field normal to the axis of rotation of the wing.
			b. Definition of stroke $\phi$ and pitch angles $\beta$ characterising the flapping wing kinematics, and c. wing dimensions.}
		\label{fig:expsetup}
	\end{figure}

	The flapping wing kinematics can be described by three angles and their temporal evolution, the stroke angle $\phi$, the pitch or rotation angle $\beta$, and the flap or elevation angle $\psi$.
	The stroke angle $\theta$ describes the position of the wing in the horizontal stroke plane (\cref{fig:expsetup}b).
	In hovering flight, the stroke follows a sinusoidal profile for most insect species.
	The pitch angle $\beta$ in \cref{fig:expsetup}b describes the rotational position of the wing and determines the geometric angle of attack.
	The pitching motion is the most complex motion function in the hovering flight kinematics and its shape varies strongly between different species \cite{ellington_aerodynamics_1984a}.
	The pitch actuation is the main focus of this study.
	The elevation angle $\psi$ is measured relative to the vector normal to the stroke plane (not shown in \cref{fig:expsetup}).
	It plays a minor role in the hovering of insects with similar Reynolds number and wing aspect ratio~\cite{liu_size_2009}.
	In this study, the flap angle is kept constant at $\psi = \ang{0}$ and the stroke angle varies sinusoidally with a fixed amplitude and frequency.
	\subsection{Dynamic scaling}
	The two non-dimensional parameters that characterise the aerodynamic properties of the flapping wing in hover are the reduced frequency $k$ and the Reynolds number $Re$.
	The reduced frequency $k$ measures the degree of unsteadiness of the flow by relating the spatial wavelength of the flow disturbance to the chord length $c$ and can be calculated as:
	\begin{equation}
	k=\frac{\pi c}{2\phi \kindex{R}{2}} \quad ,
	\label{eq:reduced_frequency}
	\end{equation}
	where $2\phi$ is the peak-to-peak stroke amplitude and $\kindex{R}{2} = \sqrt{\int_0^R(\kindex{R}{0}+r)^2 dr/R}$ is the radius to the second moment of area.
	For a rectangular wing, which is used in this study, $\kindex{R}{2}$ is also the span-wise position where the force applies~\cite{sum_wu_scaling_2019}.
	The root cutout \kindex{R}{0} of the wing is indicated in \cref{fig:expsetup}c and is the distance between the stroke axis and the wing root.

	The Reynolds number $Re$ describes the ratio between the inertial and viscous forces and is determined for the hovering flight by
	\begin{equation}
	Re=\frac{\overline{U}c}{\nu} = \frac{2\phi f c \kindex{R}{2}}{\nu} \quad ,
	\label{eq:reynolds_number}
	\end{equation}
	where $\nu$ is the kinematic viscosity of the fluid and the characteristic velocity $\overline{U} = 2\phi f R$ is defined as the stroke average wing velocity at the second moment of area $\kindex{R}{2}$~\cite{sane_control_2001, sum_wu_scaling_2019}.

	The experimental parameters for the model wing are summarised in \cref{tab:wingparam}.
They are selected to match the characteristics of larger insects, such as hawk moths, or the hummingbird in hovering flight~\cite{shyy_recent_2010, gehrke_genetic_2018}.

	The model wing used in this study has a rectangular planform (\cref{fig:expsetup}c).
Even though insect or bird wing in nature are typically not rectangular, the rectangular planform is commonly used to study the aerodynamic of revolving and flapping wings~\cite{shyy_recent_2010, jardin_three-dimensional_2012, krishna_flowfield_2018}.
The performance of flapping wings is influenced by the wing shape and geometry and the wing planform can be optimised for specific flight modes and mission profiles of the aerial vehicle~\cite{nan_experimental_2017}.
Results of systematic investigations of the influence of the wing geometry by Ansari et al.~\cite{ansari_insectlike_2008} revealed that planforms with straight leading edges are desirable for increased performance.
Therefore, we believe that the flat plate is a valid simplification of an insect wing.
Our numerical results will still be specific to the selected wing planform but the main procedure and underlying flow physics are expected to be generalisable for different more insect-inspired wing planforms.

	\begin{table}[b!]
		\centering
		\caption{Summary of the experimental parameters of the dynamically scaled wing used throughout this study. The working fluid in the experiments is water with $\kindex{\nu}{\SI{20}{\celsius}} = \SI{1.00e-06}{\meter\squared\per\second}$.}
		\label{tab:wingparam}
		\begin{tabular}{lcc}
			\hline
			Parameters & & model wing \\ \hline
			Wing stroke frequency & $f$ & \SI{0.25}{\hertz} \\
			Wing chord & $c$ & \SI{34}{\milli\meter} \\
			Wing span & $R$ &  \SI{107}{\milli\meter} \\
			Stroke amplitude & $\phi$ & \ang{180} \\
			Reduced frequency & $k$ & \num{0.19} \\
			Reynolds number & $Re$ & \num{4895} \\ \hline
		\end{tabular}
	\end{table}%
	\subsection{Experimental Setup}
	A schematic representation of the experimental setup is depicted in \cref{fig:expsetup}a.
	The flapping wing mechanism is submerged in an octagonal tank with an outer diameter of \SI{0.75}{\metre} filled with water.
	For a tip-to-tip amplitude of \SI{0.47}{\meter} this leads to a $6.91 c$ minimum tip clearance which has been shown to be sufficient to avoid wall effects in flapping wing experiments~\cite{manar_tip_2014, krishna_flowfield_2018}.
	The stroke and pitch motion are driven by two servo motors (Maxon motors, type RE35, \SI{90}{\watt}, \SI{100}{\newton\milli\meter} torque, Switzerland) reduced by $35:1$ with a planetary gear-head for the stroke and $19:1$ for the pitch actuation.
This experimental flapping wing mechanism is a unique set-up in terms of its robustness, repeatability, and the variety of kinematic motions that can be executed.
	Initial tests on the highest lift kinematics showed an error of $< \ang{0.1}$ between the motor input signal and the motor response measured by the encoder throughout the entire cycle.
	A motion controller (DMC-4040, Galil Motion Control, USA) is used to control the motors.
	The aerodynamic loads are recorded with a six-axis IP68 force-torque transducer (Nano17, ATI Industrial Automation, USA) with a resolution of \SI{3.13}{\milli\newton} for force and \SI{0.0156}{\newton\milli\meter} for torque measurements positioned at the wing root.
	The forces are recorded via a data acquisition card (National Instruments, USA) with sampling frequency of \SI{1000}{\hertz}.
	The force data was filtered with a zero phase delay low-pass \num{5}th order digital Butterworth filter.
	The cut-off frequency was chosen to be \num{12} times higher than the flapping frequency $f$.

	A high-power light-emitting diode (LED) (LED Pulsed System, ILA\_5150 GmbH, Germany) and a cylindrical lens are used to produce a \SI{4}{\milli\metre}-thick light-sheet.
	The illuminated plane of interest is recorded by a sCMOS camera (ILA\_5150 GmbH / PCO AG, Germany) with a \SI{2560x2160}{px} resolution covering a \SI{119x101}{\milli\meter} field of view.
	Phase-locked particle image velocimetry (PIV) is conducted by triggering the LED and camera simultaneously to record a single image pair for a specific phase angle $\phi$.
	To record the different phase positions throughout the stroke cycle the initial stroke angle is shifted relative to the LED-plane similar to the procedure used by Krishna et al.~\cite{krishna_flowfield_2018}.
	A total of \num{39} different stroke angle positions are recorded and averaged over \num{64} flapping cycles.
	A multi-grid algorithm with a resulting interrogation window size of \SI{48x48}{px} and an overlap of \SI{50}{\percent} is used to correlate the raw images and reconstruct the velocity flow field with a physical resolution of \SI{1.1}{\milli\meter} or \SI{0.034}{c}.
	The flow field measurements were conducted on the plane normal to the span-wise direction at the $R_2$ location or $0.56 R$ measured from the wing root (\cref{fig:expsetup}a).
	To quantify the flow properties for the converged optimisation kinematics, PIV experiments are carried out for \num{19} out of the \num{35} Pareto front individuals.

	\subsection{Optimisation}
	Genetic algorithms and other evolutionary optimisation strategies employ a survival of the fittest strategy.
	Multiple sets of parameters are tested each generation and the best performing individuals are advanced to improve further generations.
	Genetic algorithms have proven to be effective and robust for experimental data which is prone to more noise in the data.
	Due to their stochastic nature, evolutionary algorithms are strong in evading local optima which is especially important for unsteady aerodynamics where some changes in the actuation can cause a cascade of events and a drastic change in the performance.
	The objective scores of the evolutionary algorithm do not need to be weighted to be used in a multi-objective optimisation.
	This gives the genetic algorithm the natural ability to determine the trade-off between objectives in the Pareto front.

	The two optimisation targets in this study are the stroke average lift coefficient $\kindex{\overline{C}}{L}$ and the hovering efficiency $\eta$.
	The force and power coefficients of the system can be calculated from the force and torque measurements by the load transducer positioned at the root of the wing (\cref{fig:expsetup}a) according to:
	\begin{equation}
	\kindex{C}{L} = \frac{L}{\frac{1}{2}\rho R c \overline{U}^2} , \quad \kindex{C}{P} = \frac{P}{\frac{1}{2}\rho R c \overline{U}^3} \, ,
	\label{eq:aerodynamic_coefficients}
	\end{equation}
	where $L$ is the instantaneous lift, $D$ the drag, and $P$ the aerodynamic power of the system.
	For the two-axis motion, the power $P$ is calculated as the sum of pitching power $\kindex{P}{p}$ and the stroke power $\kindex{P}{s}$.
	The pitching power is the power required to rotate the wing around its pitching axis and is given by $\kindex{P}{p} = \kindex{T}{p} \dot{\beta}$, with $\kindex{T}{p}$ the measured pitch torque and $\dot{\beta}$ the angular velocity of the pitching motion.
	The stroke power is given by $\kindex{P}{s} = \kindex{T}{s} \dot{\phi}$ with $\kindex{T}{s}$ the stroke torque and $\dot{\phi}$ the stroke velocity.
	The stroke torque cannot be measured directly and is calculated from the drag force $D$ along the span $\kindex{T}{s} = \int_R D(r) r \, dr$~\cite{sane_control_2001}.
	For a uniform drag coefficient distribution along the span, the torque can be computed as $\kindex{T}{s} = D \kindex{R}{d}$, where $D$ is the drag measured at the wing root and acting on the radial position $\kindex{R}{d} = \frac{3}{4}\frac{(\kindex{R}{0}+R)^4-\kindex{R}{0}^4}{(\kindex{R}{0}+R)^3-\kindex{R}{0}^3}$.

	The hovering efficiency of the flapping wing system is computed as the ratio between the stroke average lift coefficient $\kindex{\overline{C}}{L}$ and stroke average power coefficient $\kindex{\overline{C}}{P}$:
	\begin{equation}
	\eta = \frac{\kindex{\overline{C}}{L}}{\kindex{\overline{C}}{P}} \quad .
	\label{eq:hovering_efficiency}
	\end{equation}
	This basic definition of efficiency expresses how much energy is invested to generate a certain amount of lift.
	Other definitions of the hovering efficiency quantify the dimensionless aerodynamic power to keep a unit weight in hover~\cite{wang_aerodynamic_2008} which involves specifying the weight of the hovering insect or aerial vehicle.

\begin{figure}
		\centering
		\includegraphics{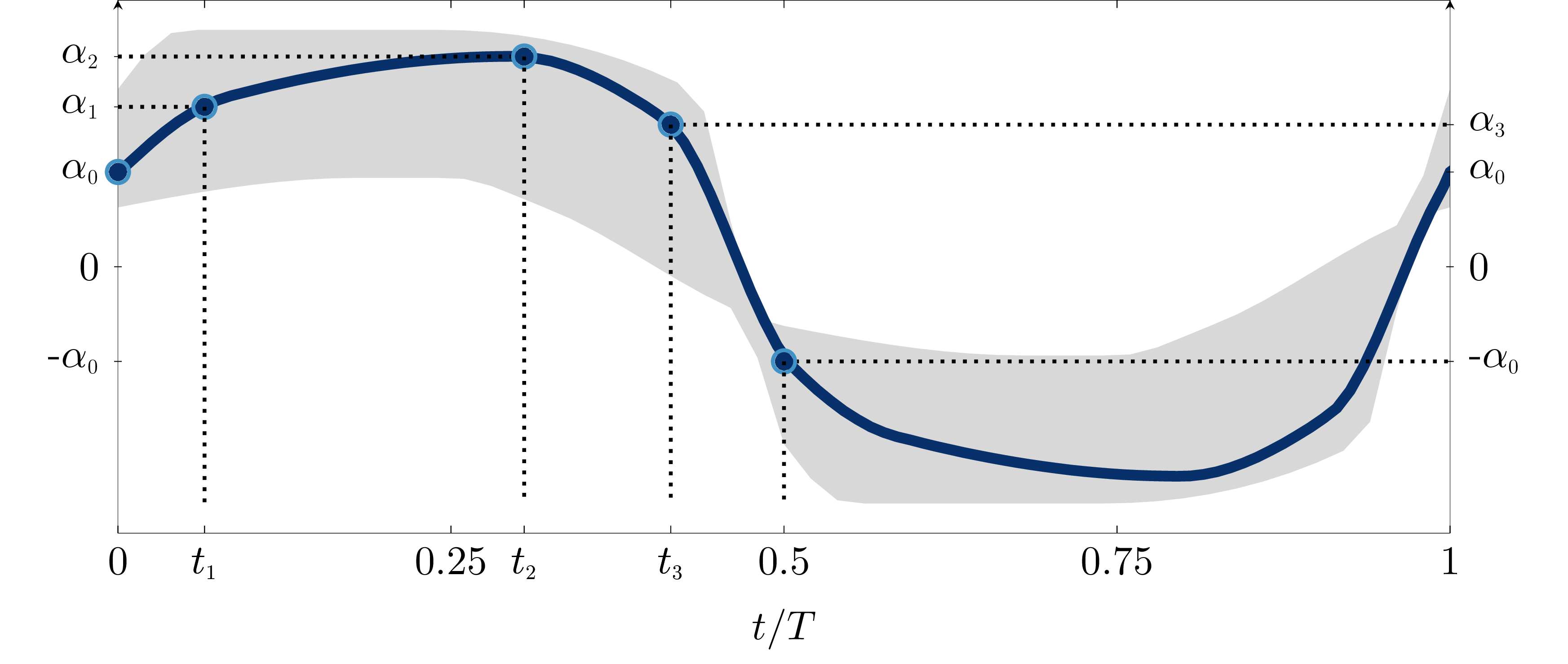}
		\caption{Pitch angle $\beta$ optimisation function used throughout the experiments.
			The four angles $\kindex{\beta}{0}, \kindex{\beta}{1}, \kindex{\beta}{2}, \kindex{\beta}{3}$ and the three phase times $\kindex{t}{1}, \kindex{t}{2}, \kindex{t}{3}$ are optimised by the evolutionary algorithm to improve the objective function.}
		\label{fig:kinematics_parametrization}
	\end{figure}

	The optimisation scheme is implemented with a genetic algorithm from the MATLAB Global optimisation Toolbox (The MathWorks Inc., USA)~\cite{chipperfield_matlab_1995}.
	Genetic algorithms explore the solution space of a process or function by using artificial evolution, a strategy also known as the survival of the fittest.
	Analogous to natural evolution, the fittest individuals of a population reproduce to ensure advancement of succeeding generations.
	In this study, seven parameters characterising the pitch angle motion $\beta$ are the genes or chromosomes in the genetic algorithm population.
	The total population consists of \num{100} individuals where the \num{35} highest performing genes make up the Pareto front individuals.
	The pitch angle function $\beta(t)$ displayed in \cref{fig:kinematics_parametrization} is defined by four parameters for the pitch angle extrema and three parameters for their respective timings.
	The parameters can vary between certain bounds listed in \cref{tab:pitchparam} to cover a wide range of possible kinematics similar to those observed in nature~\cite{liu_size_2009}.
	The objective or fitness function converts the parameters into the specific kinematics and evaluates their performance experimentally on the flapping wing system.
	Each kinematic is executed over eight consecutive flapping cycles and its fitness, the stroke average lift coefficient, and hovering efficiency, are calculated from the load cell data of the last four cycles to ensure a steady-state is reached and the influence of transient effects is limited.
 	Under certain flight conditions like forward flight or hovering with an inclined stroke planes, asymmetric stroke and pitch profiles are used~\cite{Jardin2009, park_kinematic_2012}.
	In this study, the stroke and pitch angle kinematics are symmetric and the front- and backstroke are identical which is the normal hovering flight as observed by the majority of insects~\cite{shyy_recent_2010}.
	Due to the symmetry of the prescribed motion and thanks to the high precision of our flapping wing device we obtain symmetric force and torque responses.
	The differences between the forces measurements during the front- and the backstroke are less than \SI{5}{\percent} of the maximum values.
Only the results for one half cycle are presented to give a more compact presentation of the kinematics and the aerodynamic performance.

	The initial population is randomly drawn from a uniform distribution bounded by the constraints in \cref{tab:pitchparam}.
	After all kinematics of the population have been evaluated, the individuals are ranked based on their fitness and obtain a score relative to the inverse square root of their rank.
	Several individuals of the population are then selected and their chromosomes are either used directly (cloned), randomly modified (mutated), or combined with other genes (crossover) to create the individuals for the next generation.
	The genes used for this process are chosen stochastically based on their previous performance, where a higher score leads to a higher probability to be selected.
	For the presented optimisation, \SI{5}{\percent} of the previous generation’s elite are clones, \SI{60}{\percent} are created as crossover, and \SI{35}{\percent} as mutation offsprings.
	The genes generated by the crossover function combine the parameters of two parents according to the following rule: \textrm{child = parent$_\textnormal{A}$ + rand $\times$ (parent$_\textnormal{B}$ - parent$_\textnormal{A}$)}, where \textrm{rand} is a random number between 0 and 1 drawn from a uniform distribution.
	After the new generation of offsprings is created, its fitness is evaluated by the objective function and the process continues until a predefined termination condition is reached.
	The optimisation for this study converged after \num{40} generations conducting \num{4000} experiments on the flapping wing apparatus over the course of three consecutive days.
	The evolution of the pitch angle kinematics $\beta$ progressed quickly for the first ten generations, then the solutions vary only slightly within a small margin for the remainder of the optimisation where only minor improvements are made.
	The genetic algorithm optimisation was halted after the average fitness of the Pareto front individuals did not advance within the last ten generations.
	\begin{table}[tb]
		\centering
		\caption{Parameter bounds for the pitching motion optimisation}
		\label{tab:pitchparam}
		\begin{tabular}{lccccccc}
			\hline
			& $\kindex{\beta}{0}$ & $\kindex{\beta}{1}$ & $\kindex{\beta}{2}$ & $\kindex{\beta}{3}$ & $\kindex{t}{1}$ & $\kindex{t}{2}$ & $\kindex{t}{3}$ \\ \hline
			minimum & $\ang{30}$ & $\ang{30}$ & $\ang{30}$ & $\ang{20}$ & $0.05T$ & $\kindex{t}{1}+0.2(\kindex{t}{3}-\kindex{t}{1})$ & $0.33T$ \\
			maximum & $\ang{60}$ & $\ang{75}$ & $\ang{75}$ & $\ang{60}$ & $0.18T$ & $\kindex{t}{3}-0.2(\kindex{t}{3}-\kindex{t}{1})$ & $0.43T$ \\ \hline
		\end{tabular}%
	\end{table}%
	\section{Results}%
	\subsection{Phenomenological Overview} \label{sec:phenomenological_overview}
\begin{figure}[tb!]
		\centering
		\includegraphics[]{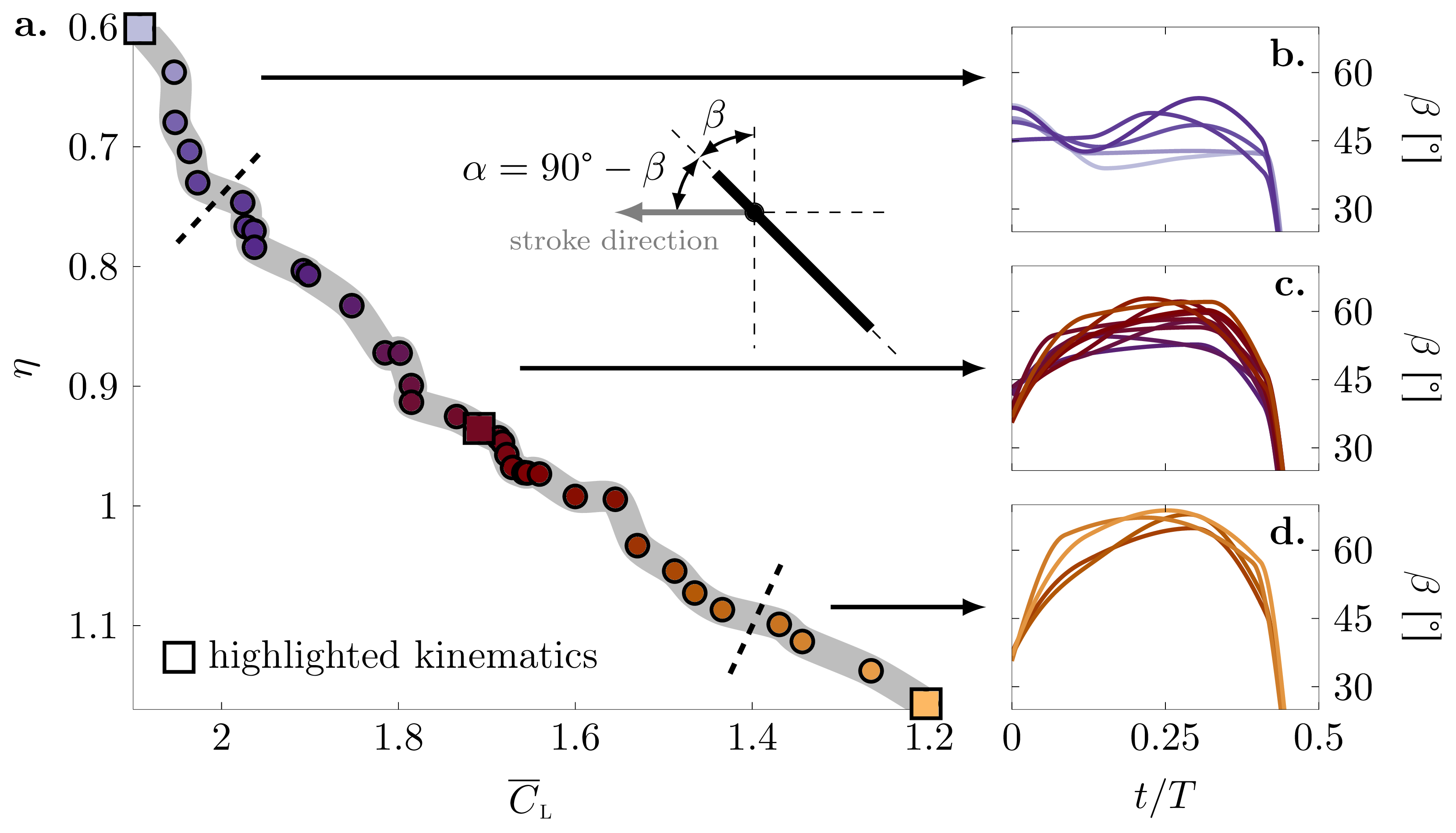}
		\caption{a.\@ Final Pareto front for the optimisation objectives hovering efficiency $\eta$ vs stroke average lift coefficient $\kindex{\overline{C}}{L}$.
			Solutions marked with a square marker are examined in more detail in Section~\ref{sec:phenomenological_overview}\@.
			b.\@-d.\@ Temporal evolution of the pitch angle $\beta$ for a single stroke for different sections of the Pareto front.}
		\label{fig:pareto_front}
	\end{figure}

	The two optimisation objectives in this study are the stroke average lift coefficient $\kindex{\overline{C}}{L}$ and the stroke average hovering efficiency $\eta$.
	The final shape of the Pareto front in \cref{fig:pareto_front}a represents the trade-off between those two optimisation targets.
	The coloured markers represent the individuals of the final generation on the Pareto front whose specific kinematics and associated aerodynamic loads will be analysed in more detail here.
	The x and y-axis in \cref{fig:pareto_front}a have been inverted following standard conventions.
	The stroke-average lift $\kindex{\overline{C}}{L}$ produced by the optimised kinematics ranges from \numrange{1.20}{2.09} and the aerodynamic performance $\eta$ varies from \numrange{0.60}{1.17}.
	By trading off up to \SI{43}{\percent} of its maximum lift capacity, the flapping wing system's efficiency can be increased by \SI{93}{\percent} by merely adjusting the pitch angle kinematics.

	The Pareto front can be divided in three sections based on the local change in the gradient $\textrm{d}\eta/\textrm{d}\kindex{\overline{C}}{L}$ along the front.
	Along the large central part of the Pareto front, the lift increases approximately linearly with decreasing efficiency.
	In this bulk part of the Pareto front, an increase of $\Delta \kindex{\overline{C}}{L}=0.1$ costs $\Delta\eta=0.058$ or an increase of $\Delta\eta=0.1$ costs $\Delta \kindex{\overline{C}}{L}=0.167$.
	Near the tails of the Pareto front, there is a larger trade off between lift and efficiency.
	For the highest lift cases, we can squeeze out an increase of $\Delta \kindex{\overline{C}}{L}=0.1$ at the expense of losing $\Delta\eta=0.138$.
	For the highest efficiency cases, we can squeeze out an increase of $\Delta\eta=0.1$ at the expense of losing $\Delta \kindex{\overline{C}}{L}=0.257$.


	The pitch angle kinematics $\beta$ have a distinctly different evolution for the three different regions of the Pareto front.
	The evolutions of $\beta$ are presented in \cref{fig:pareto_front}b-d for half of the flapping cycle.
	The motion is perfectly symmetric and the front- and backstroke are identical.
	The selected axes limits highlight the variations of $\beta$ during the main portion of the stroke prior to the rapid end of stroke rotation where $\beta$ drops to zero for all kinematics.
	All kinematics on the Pareto front have an advanced rotation, which means that the majority of the end of stroke rotation occurs before the end of the stroke.
	The pitch angle is the function optimised by the genetic algorithm.
	The aerodynamic angle of attack $\alpha$ during this stroke is related to the pitch angle as $\alpha=\ang{90}-\beta$.

	The kinematics in the high lift tail of the Pareto front (\cref{fig:pareto_front}b) have an almost trapezoidal pitch angle profile.
	The pitch angle is more or less constant around $\beta=\ang{45}$ for $0<t/T<0.4$ and there is an abrupt end of stroke rotation.
	The kinematics in the high efficiency tail of the Pareto front (\cref{fig:pareto_front}d) have a more rounded, sinusoidal profile with a maximum pitch angle $\beta>\ang{60}$ around mid-stroke.
	The high pitch angle leads to a substantially lower angle of attack in the high efficiency tail compared to the high lift tail.
	The transition into the end of stroke rotation is smooth.
	The kinematics in the bulk of the Pareto front (\cref{fig:pareto_front}c) gradually evolve from the more trapezoidal high lift kinematics towards the almost sinusoidal high efficient kinematics.

	The intermediate and most efficient pitch angle kinematics obtained in the optimisation resemble pitch angle evolutions observed in a dynamically scaled crane fly model in hover~\cite{ishihara_experimental_2014}.
The stroke actuation of this crane fly model was fixed while the wing was allowed to passively pitch in response to the aerodynamic forces.
	Similar pitch angle profiles were also found as efficient hovering motions of a hawkmoth obtained by a numerical-based optimisation~\cite{lee_optimization_2018}.
	The high lift kinematics along the Pareto front share the same features that can be observed in the free-hovering flight wing kinematics of a horned beetle~\cite{phan_wing_2018}.
The wings of the horned beetle are flexible and significant contributions of the wing inertia and elastic storage from the wing deformation lower the total power requirements for the hovering motion.
Yet, the pitch kinematics found for these natural flexible wings with different planform shapes match the solutions obtained by our optimisation.
This confirms that we are able to optimise the underlying aerodynamic effects that govern effective flapping wing flight with the use of a rectangular rigid plate.

	To understand and characterise the variations between the different kinematic solutions and their force and flow responses, we have selected three solutions along the Pareto front to guide the description.
	The selected solutions are the highest lift generating, the most efficient, and an intermediate solution.
	They are indicated by the square markers in \cref{fig:pareto_front}a.

\begin{figure}[tb!]
		\centering
		\includegraphics[]{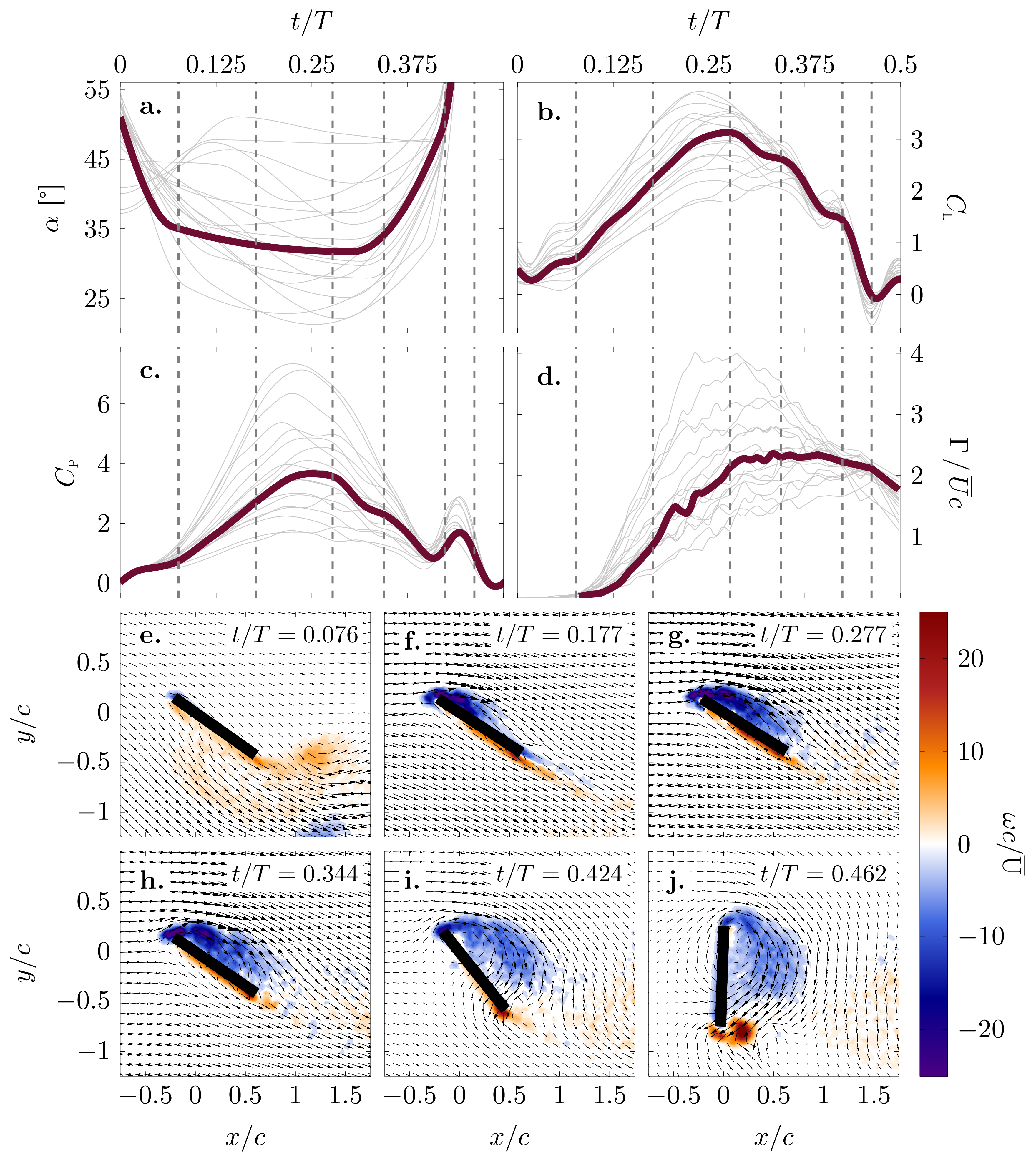}
		\caption{Overview of the intermediate Pareto front kinematics and their aerodynamic performance ($\kindex{\overline{C}}{L} = 1.71, \eta = 0.94$).
			a.~Temporal evolution of the angle of attack $\alpha$, b.~lift coefficient $\kindex{\overline{C}}{L}$, c.~power coefficient $\kindex{C}{P}$, and d. leading edge vortex circulation $\Gamma$.
			e.-j.~Selected velocity and vorticity fields within a single stroke.
				The grey lines in a.-d. represent all Pareto optimal solutions for reference.
				The bold lines represent the results for the intermediate Pareto front kinematics.
		}
		\label{fig:forces_PIV_intermediate}
	\end{figure}

\begin{figure}[tb!]
		\centering
		\includegraphics[]{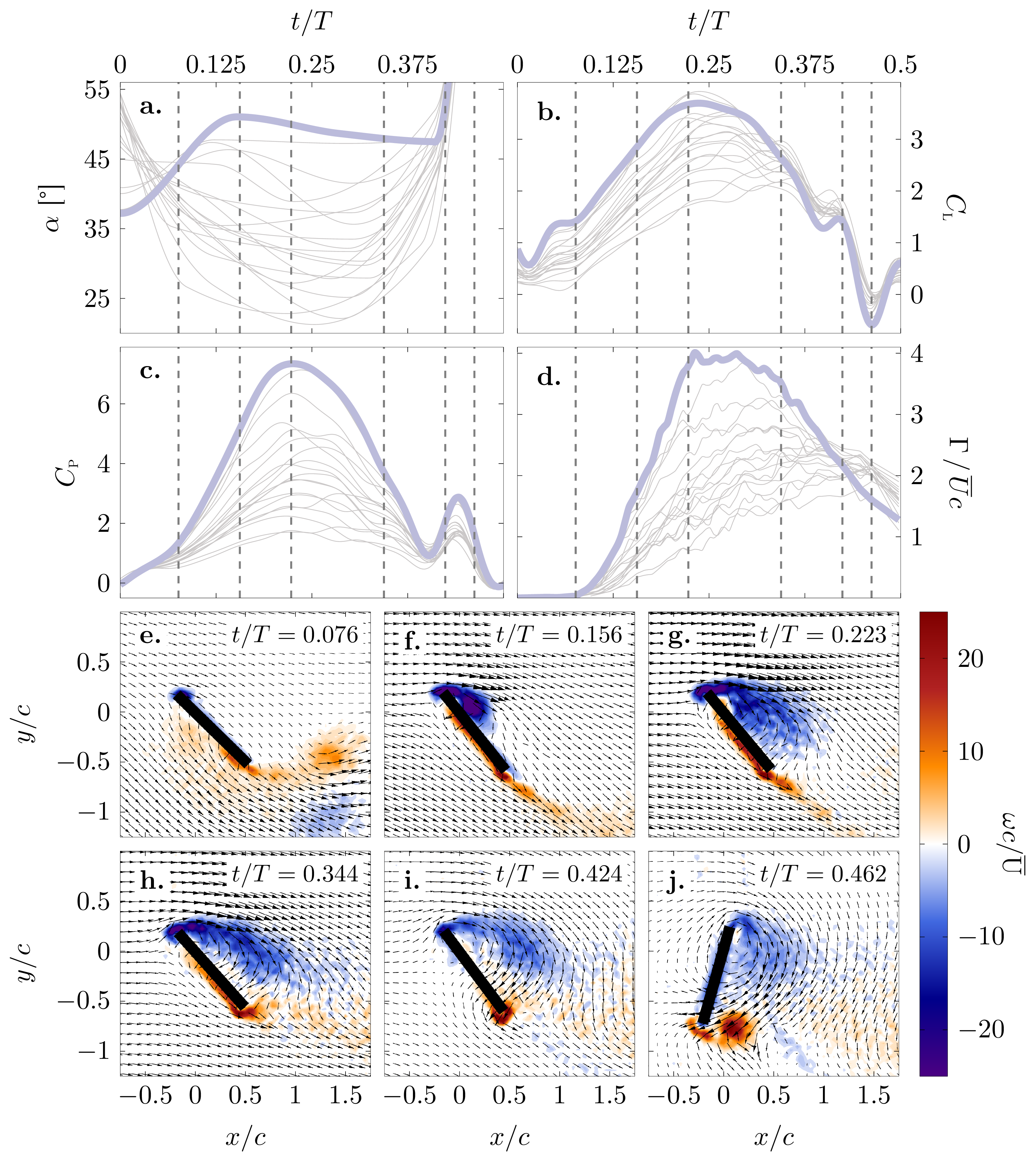}
		\caption{Overview of the highest lift generating Pareto front kinematics and their aerodynamic performance ($\kindex{\overline{C}}{L} = 2.09, \eta = 0.60$).
			a.~Temporal evolution of the angle of attack $\alpha$, b.~lift coefficient $\kindex{\overline{C}}{L}$, c.~power coefficient $\kindex{C}{P}$, and d. leading edge vortex circulation $\Gamma$.
			e.-j.~Selected velocity and vorticity fields within a single stroke.
			The grey lines in a.-d. represent all Pareto optimal solutions for reference.
			The bold lines represent the results for the highest lift generating Pareto front kinematics.
		}
		\label{fig:forces_PIV_highCL}
	\end{figure}

\begin{figure}[tb!]
		\centering
		\includegraphics[]{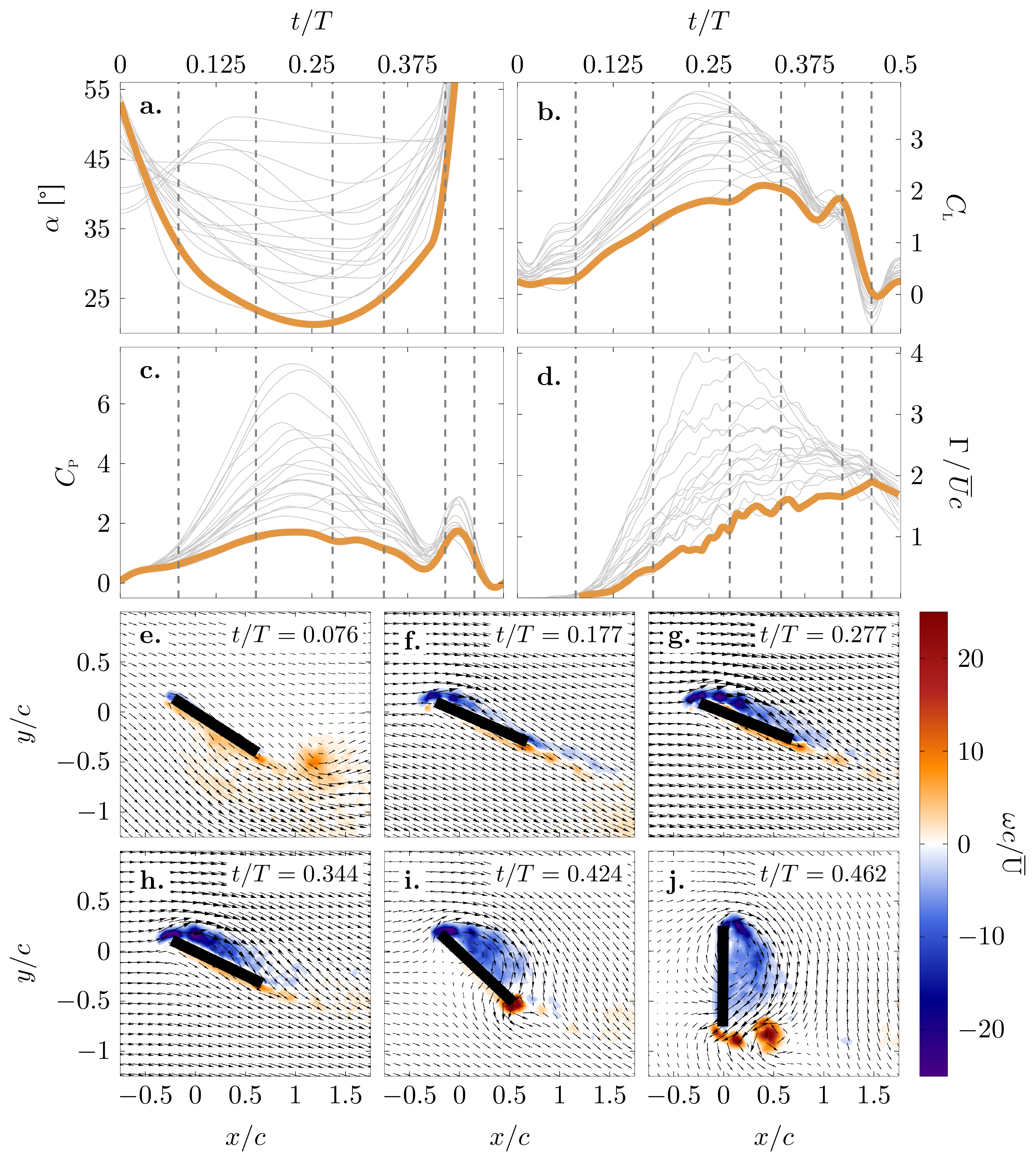}
		\caption{Overview of the most efficient Pareto front kinematics and their aerodynamic performance ($\kindex{C}{L} = 1.20, \eta = 1.17$).
			a.~Temporal evolution of the angle of attack $\alpha$, b.~lift coefficient $\kindex{\overline{C}}{L}$, c.~power coefficient $\kindex{C}{P}$, and d. leading edge vortex circulation $\Gamma$.
			e.-j.~Selected velocity and vorticity fields within a single stroke.
			The grey lines in a.-d. represent all Pareto optimal solutions for reference.
			The bold lines represent the results for the most efficient Pareto front kinematics.
		}
		\label{fig:forces_PIV_lowCL}
	\end{figure}

	The pitching kinematics of the selected cases and their flow and forces responses are presented first for the intermediate solution (\cref{fig:forces_PIV_intermediate}), then for the highest lift generating (\cref{fig:forces_PIV_highCL}), and finally for the most efficient solution (\cref{fig:forces_PIV_lowCL}).
	The pitching kinematics are expressed now in terms of the aerodynamic angle of attack $\alpha$.
	Only one half cycle is shown in the figures for a more concise representation of the results.
	Both the kinematics and the corresponding force responses are symmetric between front- and backstroke.
The differences between the forces measurements during the front- and the backstroke are less than \SI{5}{\percent} of the maximum values which is of the same order of magnitude as the differences between cycles.
	The flow and force responses are summarised by selected snapshot of the velocity and vorticity field and the temporal evolutions of the lift and power coefficient and the leading edge vortex circulation.
	The leading edge vortex circulation was computed inside the $\Gamma_2$-contour with $\Gamma_2 = 0.5$ and a radius of 5 pixels over one half-cycle~\cite{graftieaux_combining_2001}.



	The summary of the input kinematics and their response for the intermediate solution along the Pareto front corresponding to $\kindex{\overline{C}}{L} = 1.71$ and $\eta = 0.94$ is presented in \cref{fig:forces_PIV_intermediate}.
	The angle of attack of the intermediate kinematics in \cref{fig:forces_PIV_intermediate}a is already reduced to \ang{51} at the start of the stroke due to the advanced end-of-stroke rotation.
	Initially, the angle of attack continues to decrease rapidly while the stroke velocity increases.
	When the angle of attack has reached a value of \ang{35} around $t/T=\num{0.06}$, a leading edge vortex starts to form (\cref{fig:forces_PIV_intermediate}e).
	The wing in the flow field snapshots accelerates from right to left.
	While the leading edge vortex grows in chord-wise direction, the angle of attack continues to decrease but at a lower rate than before.
	Despite the gradual decrease of the angle of attack, the lift and power coefficients increase in the first half of the stroke.
	The increase is due to the growing leading edge vortex (\cref{fig:forces_PIV_intermediate}e-g) and the influence of the sinusoidal stroke motion.
	The power coefficient reaches a maximum value around the mid-stroke at $t/T=\num{0.25}$.
	The lift coefficient reaches a maximum value shortly thereafter around $t/T=\num{0.28}$ when the leading edge vortex circulation levels off.
	At this point, leading edge vorticity covers the entire chord length and continues to spread in the chord-normal direction (\cref{fig:forces_PIV_intermediate}g-h).
	This stage in the vortex development is know as vortex lift-off \cite{krishna_flowfield_2018}.
	Once the vortex lifts off of the wing, its circulation no longer increases and the lift decreases.
	Around $t/T=\num{0.33}$, the wing starts its end-of-stroke rotation and the angle of attack increases rapidly when the wing rotates back to its vertical orientation.
	The axes limits in \cref{fig:forces_PIV_intermediate}a highlight the variations in the angle of attack during the main portion of the stroke and cut off the fast rotation at the end of the stroke.
	The fast end-of-stroke pitch rotation pushes the leading edge vortex away and prompts the formation of a trailing edge vortex (\cref{fig:forces_PIV_intermediate}i-j), which yields a secondary peak in the power coefficient (\cref{fig:forces_PIV_intermediate}c).

\begin{figure}[tb!]
		\centering
		\includegraphics[]{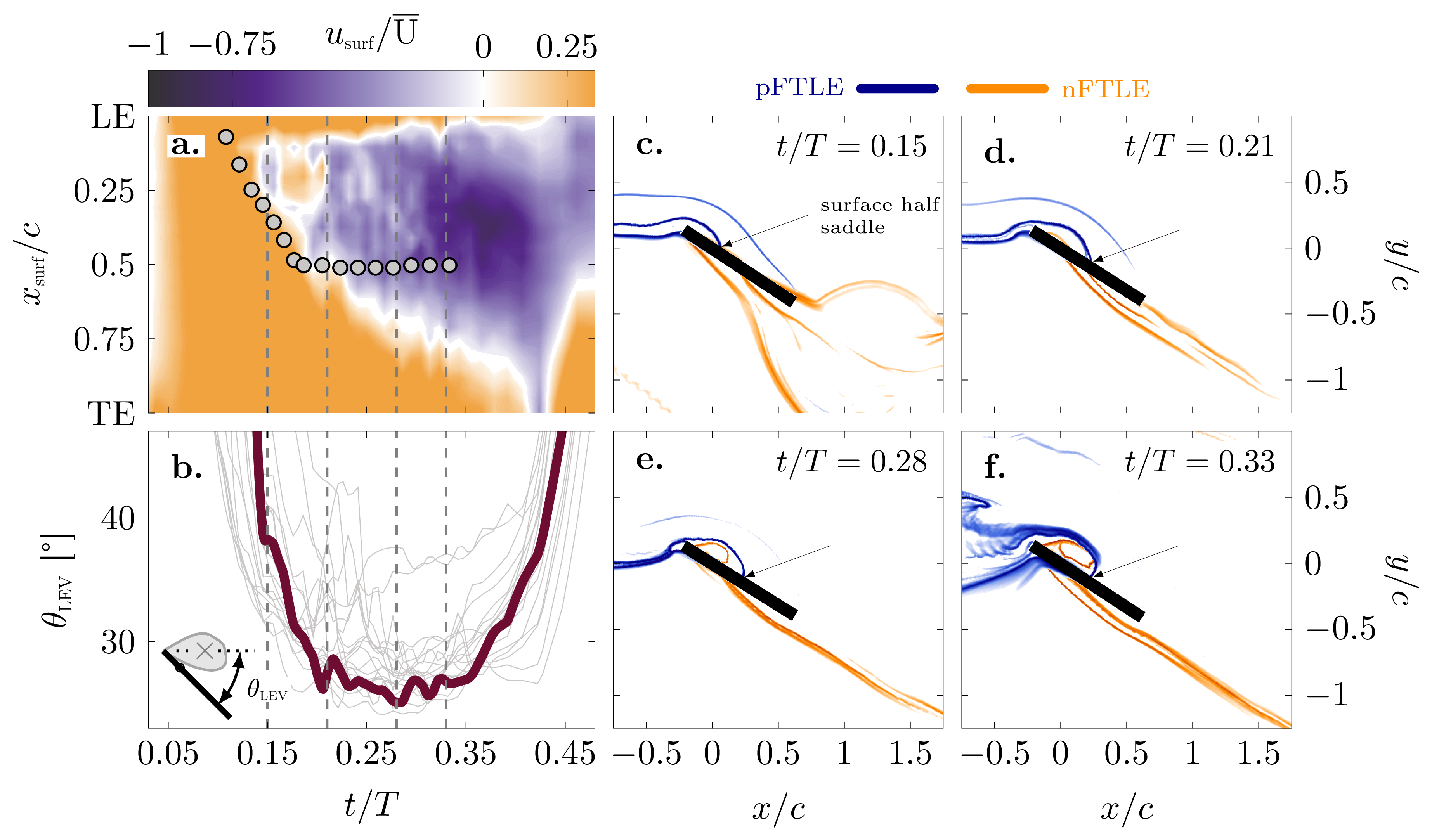}
		\caption{
			a.~Space-time representation of the surface velocity.
			The grey dots indicate the trajectory of the surface half saddle.
			b.~evolution of the angular location of the leading edge vortex with respect to the wing and the leading edge.
The grey lines in b represent all Pareto optimal solutions as reference.
The bold line represents the intermediate Pareto front kinematics ($\kindex{\overline{C}}{L} = 1.71, \eta = 0.94$).
			c.-f.~Snapshots of the FTLE ridges for selected time instants indicated by the vertical dashed lines in a and b.
		}
		\label{fig:surfvelo_intermediate}
	\end{figure}

	The different phases of the leading edge vortex development are more clearly visualised in \cref{fig:surfvelo_intermediate} by the space-time representation of the surface velocity, snapshots of the finite time Lyapunov exponent (FTLE) ridges, and the position of the leading edge vortex with respect to the wing.
	The leading edge vortex position is determined as the vorticity density centre within the $\Gamma_2=0.5$-contour.
	\Cref{fig:surfvelo_intermediate}a shows the spatiotemporal evolution of the velocity component $\kindex{u}{surf}$ parallel to the wing's surface, close to the surface.
	Positive values of $\kindex{u}{surf}$ indicate a surface flow towards the trailing edge indicative of attached surface parallel flow.
	Negative values of $\kindex{u}{surf}$ indicate a surface flow towards the leading edge induced by a leading edge vortex.
	From $t/T=$ \numrange{0.13}{0.21}, the leading edge vortex emerges at the leading edge and gradually spreads in the chord-wise direction but never fully covers the wing chord.
	This is clearly visualised by the region of negative surface velocity which gradually grows towards the trailing but only covers about \SI{75}{\percent} of the chord at $t/T\approx\num{0.33}$ when the end of stroke motions sets in.
	The limited chordwise growth of the leading edge vortex is also evidenced by the expansion of the positive-time FTLE ridges that indicate the outer boundary of the vortex.
	The FTLE fields are calculated from the phases averaged velocity fields following the same procedure as described by Krishna et al.~\cite{krishna_flowfield_2018,Krishna_2019}.
	The scalar FTLE field is a measure of local Lagrangian stretching of nearby trajectories as the flow evolves in space and time.
	The stretching of particle trajectories can can be calculated forward and backward in time to yield positive and negative-time FTLE fields (pFTLE and nFTLE).
	The maximising ridges of the FTLE fields are effective at identifying coherent structure boundaries and aid to analyse the dynamics in vortex-dominated flows \cite{Huang_Green_2015, Rockwood_Huang_Green_2018}.
	The ridges in the nFTLE fields indicate candidate attracting material lines along which particle trajectories will locally contract.
	The ridges in the pFTLE fields indicate candidate repelling material lines along which particle trajectories will diverge.
	The points along the chord where pFTLE ridges seem to meet the wing surface downstream of the leading edge vortex mark the location of surface half saddle points.
	The location of a half saddle point indicates the extend of the vortex.
	This surface half saddle moves downstream in time while the region of negative surface vorticity grow until reaching mid-chord at $t/T\approx\num{0.2}$.
	Hereafter, the surface half saddle does not move further downstream and the leading edge vortex grows in chord-normal direction.
	The downstream trajectory of the surface half saddle is added on top of the surface velocity in \cref{fig:surfvelo_intermediate}a.
	The end of the chord-wise growth coincides with the saturation of the leading edge vortex circulation.
	The end of the chord-wise vortex growth can also be observed by analysing the evolution of the angular position of the leading edge vortex with respect to the wing's surface in \cref{fig:surfvelo_intermediate}b.
	The angle \kindex{\theta}{LEV} is defined as the angle between the wing's surface and the line connection the leading edge and the vorticity density centre marking the position of the leading edge vortex as indicate in the sketch in \cref{fig:surfvelo_intermediate}b.
	This angle can also be interpreted as the angle of the shear layer that feeds the leading edge vortex.
	When the vortex grows in chord-wise direction, \kindex{\theta}{LEV} decreases rapidly until reaching a local minimum values of about \ang{25} at $t/T\approx\num{0.20}$.
	Hereafter, the angle remains approximately constant and increases again once the end-of-stroke rotation has set in.
	The apparent stagnation of the surface half saddle and the angle \kindex{\theta}{LEV} between $t/T=\num{0.2}$ and $t/T=\num{0.33}$ indicates that the leading edge vortex remains stable without growing in size and circulation but not not shedding into the wake either.


	The flow and force response for the highest lift generating kinematics along the Pareto front are presented in \cref{fig:forces_PIV_highCL}.
	The angle of attack for the intermediate kinematics started around $\ang{51}$ and decreased to $\alpha=\ang{35}$ where it remained for the majority of the stroke.
	The evolution of the angle of attack for the highest lift generating cases is the other way around.
	The angle is slightly above \ang{35} at the start of the cycle and increases to values around $\alpha=\ang{50}$ until a very abrupt end of stroke motion sets in.
	The overall higher angles of attack lead to higher values of lift, power, and leading edge vortex circulation during the entire cycle.
	The leading edge vortex development (\cref{fig:forces_PIV_highCL}e-j) is similar to the intermediate case, but the vortex evolves faster due to the higher angle of attack and associated higher circulation rate.
	This also results in earlier achievement of the maximum lift, power, and circulation.
	The higher lift generating kinematics thus prefer a higher overall angle of attack yielding a stronger leading edge vortex that reaches its maximum capacity earlier.

\begin{figure}[tb!]
		\centering
		\includegraphics[]{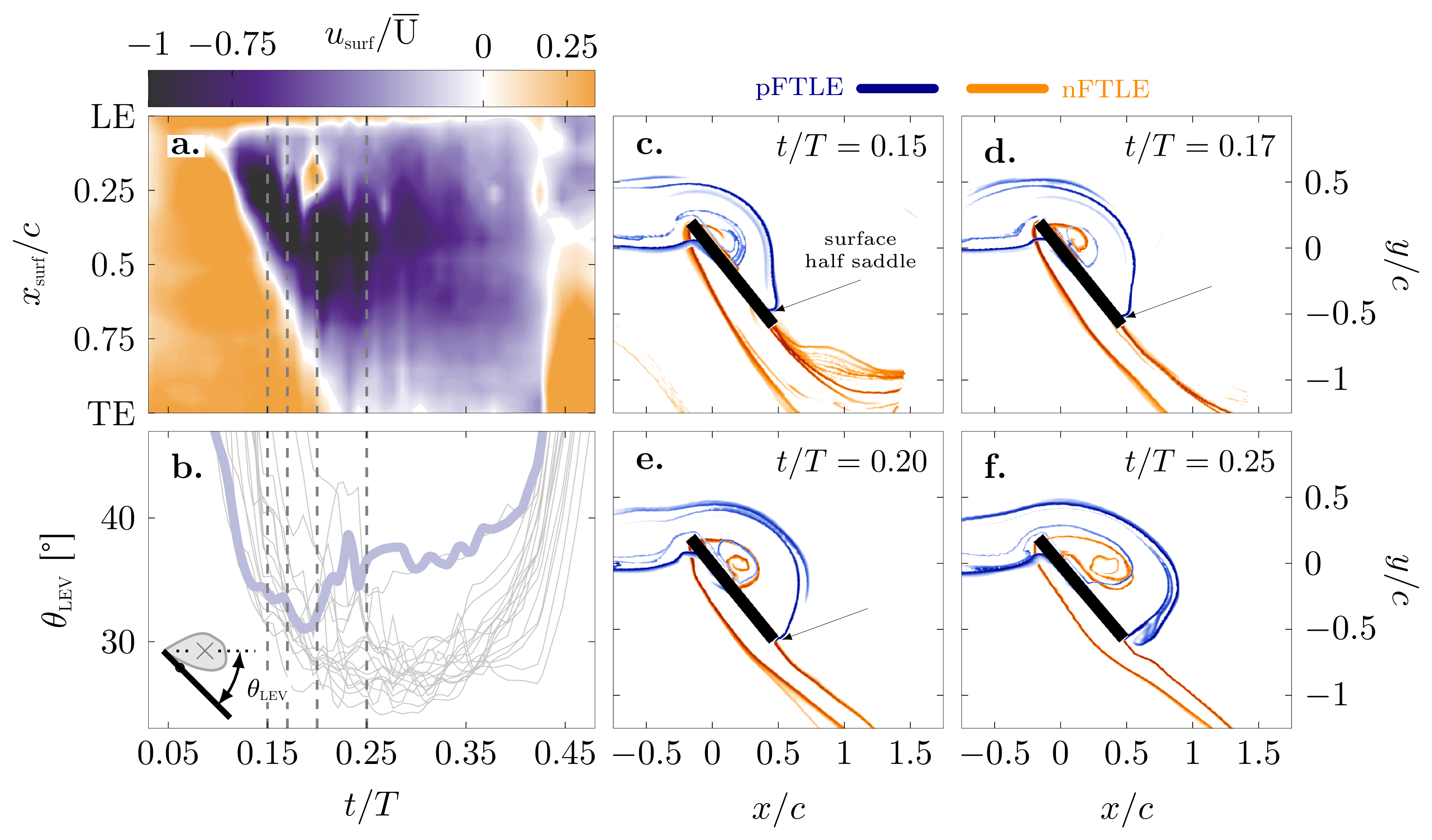}
		\caption{a.~Space-time representation of the surface velocity,
			b.~evolution of the angular location of the leading edge vortex with respect to the wing and the leading edge.
			The grey lines in b represent all Pareto optimal solutions as reference.
 			The bold line represents the highest lift generating Pareto front kinematics ($\kindex{\overline{C}}{L} = 2.09, \eta = 0.60$).
			c.-f.~Snapshots of the FTLE ridges for selected time instants indicated by the vertical dashed lines in a and b.
		}
		\label{fig:surfvelo_highCL}
	\end{figure}

	The space-time representation of the surface velocity in \cref{fig:surfvelo_highCL}a confirms the faster evolution of the leading edge vortex.
	The negative surface velocity starts to spread earlier and reaches all the way through the trailing edge by $t/T\approx{0.20}$.
	For the intermediate case, this did not occur prior to the end-of-stroke rotation.
	This moment coincides with the moment the surface half saddle point extracted from the pFTLE ridges reaches the trailing edge (\cref{fig:surfvelo_highCL}c-e).
	When the surface half saddle reaches the trailing edge, it merges with the half saddle at the trailing edge stagnation point into a full saddle that will move away from the wing marking the separation or the lift off of the vortex \cite{rival_characteristic_2014, Huang_Green_2015, krishna_flowfield_2018}.
	The time at which the vortex can no longer grow in the chordwise direction again coincides with the moment the angle $\kindex{\theta}{LEV}$ reaches a minimum value (\cref{fig:surfvelo_highCL}b).
	The local minimum in \kindex{\theta}{LEV} and the surface half saddle and negative surface velocity reaching the trailing edge all indicate the end of the growth of the leading edge vortex.
	The end of the vortex growth is followed by vortex lift off.
	The vortex lift off is significantly faster and more pronounced than for the intermediate case and all other cases which are shown by the thin grey lines (\cref{fig:surfvelo_highCL}b).
	The minimum value of the angle $\kindex{\theta}{LEV}$ is higher which indicates that the vortex is less shielded by the wing, which explains the higher drag and higher power coefficient that in required to execute these kinematics.
	The earlier vortex lift-off gives also more opportunity for a trailing edge vortex to roll up around the trailing edge.
	This leads to a higher secondary peak in the power coefficient (\cref{fig:forces_PIV_highCL}i-j).


	The flow and force response for the most efficient kinematics along the Pareto front are presented in \cref{fig:forces_PIV_lowCL}.
	The evolution of the angle of attack (\cref{fig:forces_PIV_lowCL}a) varies more gradually than all other kinematics and reaches values as low as \ang{21} around mid-stroke.
	At these low angles of attack, the leading edge vorticity remains close to the wing's surface and the circulation continues to grow during the entire stroke until the end-of-stroke motion sets in.
	The low angles of attack and compact distribution of the vorticity close to the wing lead to low values of the power coefficient during the entire stroke.
	The largest power values are now observed during the end-of-stroke rotation.
	The overall lift coefficient is also reduced as a result of the low angles and the lower vortex circulation, but it continues to increase during most of the stroke.

\begin{figure}[tb!]
		\centering
		\includegraphics[]{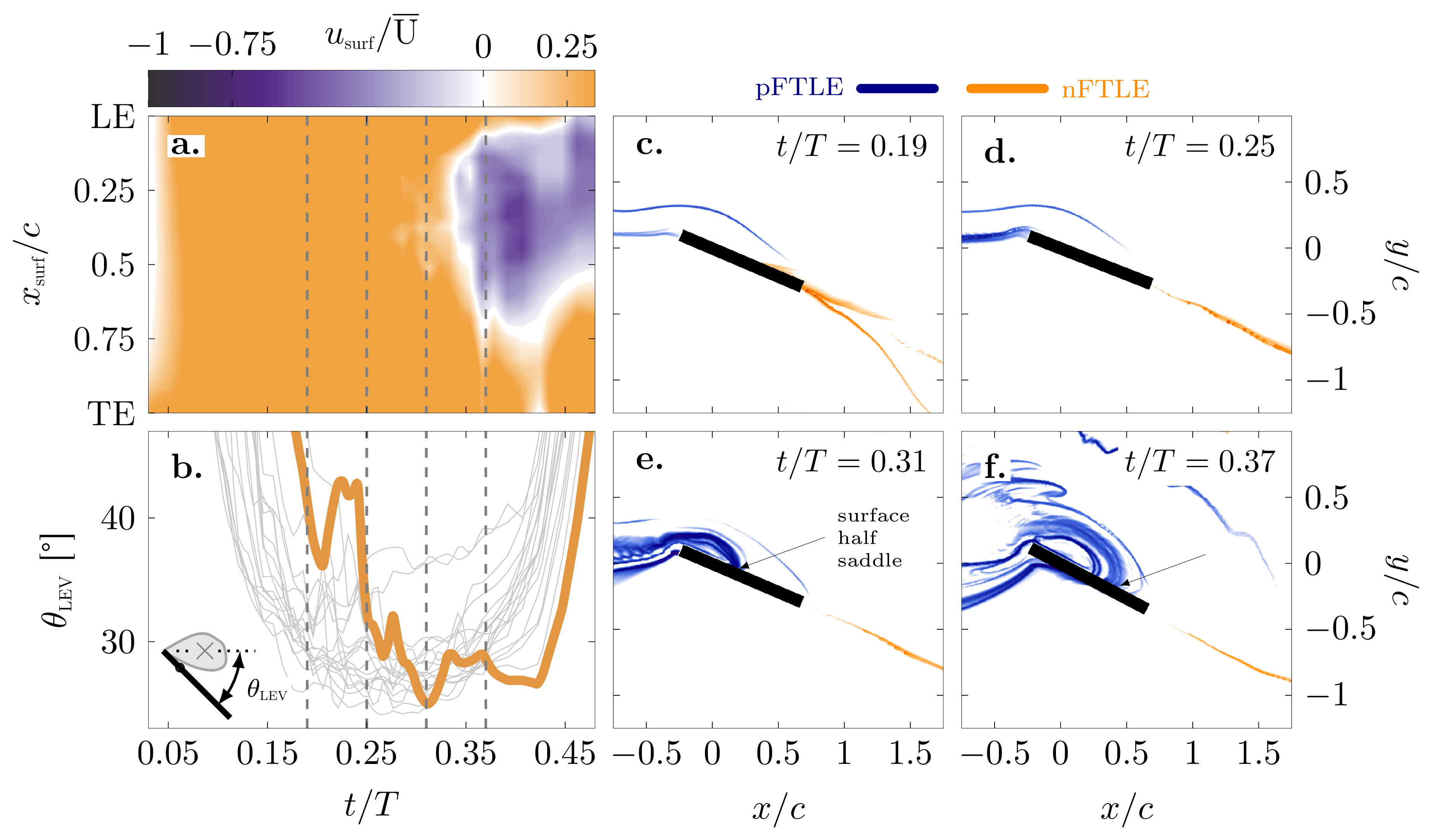}
		\caption{a.~Space-time representation of the surface velocity,
			b.~evolution of the angular location of the leading edge vortex with respect to the wing and the leading edge.
The grey lines in b represent all Pareto optimal solutions as reference.
The bold line represents the most efficient Pareto front kinematics ($\kindex{C}{L} = 1.20, \eta = 1.17$).
	c.-f.~Snapshots of the FTLE ridges for selected time instants indicated by the vertical dashed lines in a and b.
		}
		\label{fig:surfvelo_lowCL}
	\end{figure}

	The high lift generating kinematics aimed to accelerate the leading edge vortex development to create a larger and stronger vortex around mid-stroke.
	The most efficient kinematics seem to be doing the opposite and slowing down the vortex growth to delay vortex lift-off and reduce the power by keeping the vortex bound to the wing.
	This is confirmed by the surface velocity, FTLE saddle points, and the evolution of $\kindex{\theta}{LEV}$ in \cref{fig:surfvelo_lowCL}.
	The negative surface velocity spreads slower than in the previous cases and does not cover the entire surface before the end-of-stroke rotation sets in (\cref{fig:surfvelo_lowCL}a).
	The surface half saddles also do not reach the trailing edge and do not lift off (\cref{fig:surfvelo_lowCL}c-f).
	Due to the close proximity of the leading edge vorticity to the wing, the calculation of the vortex position is more sensitive and the evolution of $\kindex{\theta}{LEV}$ is a little more noisy.
	Yet, the angle does not really start to increase before the end-of-stroke rotation confirming the absence of vortex lift off and the associated penalty on the power coefficient.


	\subsection{Quantitative analysis and scaling}

	In the previous section, we qualitatively linked three characteristic flapping wing pitch angle kinematics along the Pareto front to their aerodynamic response based on the spatiotemporal evolution of the leading edge vortices that are created.
	In the reminder of the paper, we aim to quantitatively describe and scale the temporal evolution of the vortex development and the aerodynamic forces and efficiency for all solutions along the Pareto front.
	First, we will extract characteristic velocity and time scales directly from the kinematic input.
	Second, we will demonstrate how these characteristic parameters allow us to scale the leading edge vortex circulation and the aerodynamic performance.

	The leading edge vortex formation on plunging and translating plates rapidly accelerating from rest is well described based on the effective velocity of the leading edge shear layer~\cite{kriegseis_persistence_2013}.
	The effective shear layer velocity for the leading edge vortex formation on pitching and rotating flat plates can be approximated by the leading-edge-normal velocity due to the motion kinematics~\cite{manar_comparison_2016}.
	For our flapping wing hovering motion, the time dependent shear layer velocity \kindex{u}{s} at the span-wise location corresponding to the second moment of area of the wing ($R_2$) is calculated using the stroke velocity $\dot{\phi}$ and pitch velocity $\dot{\beta}$ components:
	\begin{equation}
	u_s(t) = R_2 \, \dot{\phi}(t) \cos\left(\beta(t)\right) + 0.25 c \, \dot{\beta}(t)\quad.
	\label{eq:inst_shear_layer_velo}
	\end{equation}
	The second moment of area of the wing $R_2$ is the span-wise location where the force is applied~\cite{sum_wu_scaling_2019} and the evolution of the shear layer velocity at this location serves as representative velocity for the analysis of the vortex dynamics.
	By integrating the temporal evolution of the leading edge shear layer velocity \kindex{u}{s} defined by \cref{eq:inst_shear_layer_velo} over a time $t$ since the start of the flapping cycle, we obtain the advective time $\sigma$:
	\begin{equation}
	\sigma(t) = \int_{0}^{t} \kindex{u}{s}(\tau) d\tau = \int_{0}^{t} R_2 \, \dot{\phi}(\tau) \cos\left(\beta(\tau)\right) + 0.25 c \, \dot{\beta}(\tau) d\tau\quad.
	\label{eq:adv_time1}
	\end{equation}
	The advective time describes the distance the leading edge has traveled since the beginning of the stroke-cycle~\cite{manar_comparison_2016}.
	The shear layer velocity is a measure for the instantaneous feeding rate of vorticity into the leading edge vortex.
	The advective time is a measure for the total amount of vorticity fed into the vortex since the start of the stroke motion and indicates the age of the vortex.

\begin{figure}[tb]
		\centering
		\includegraphics[width=\textwidth]{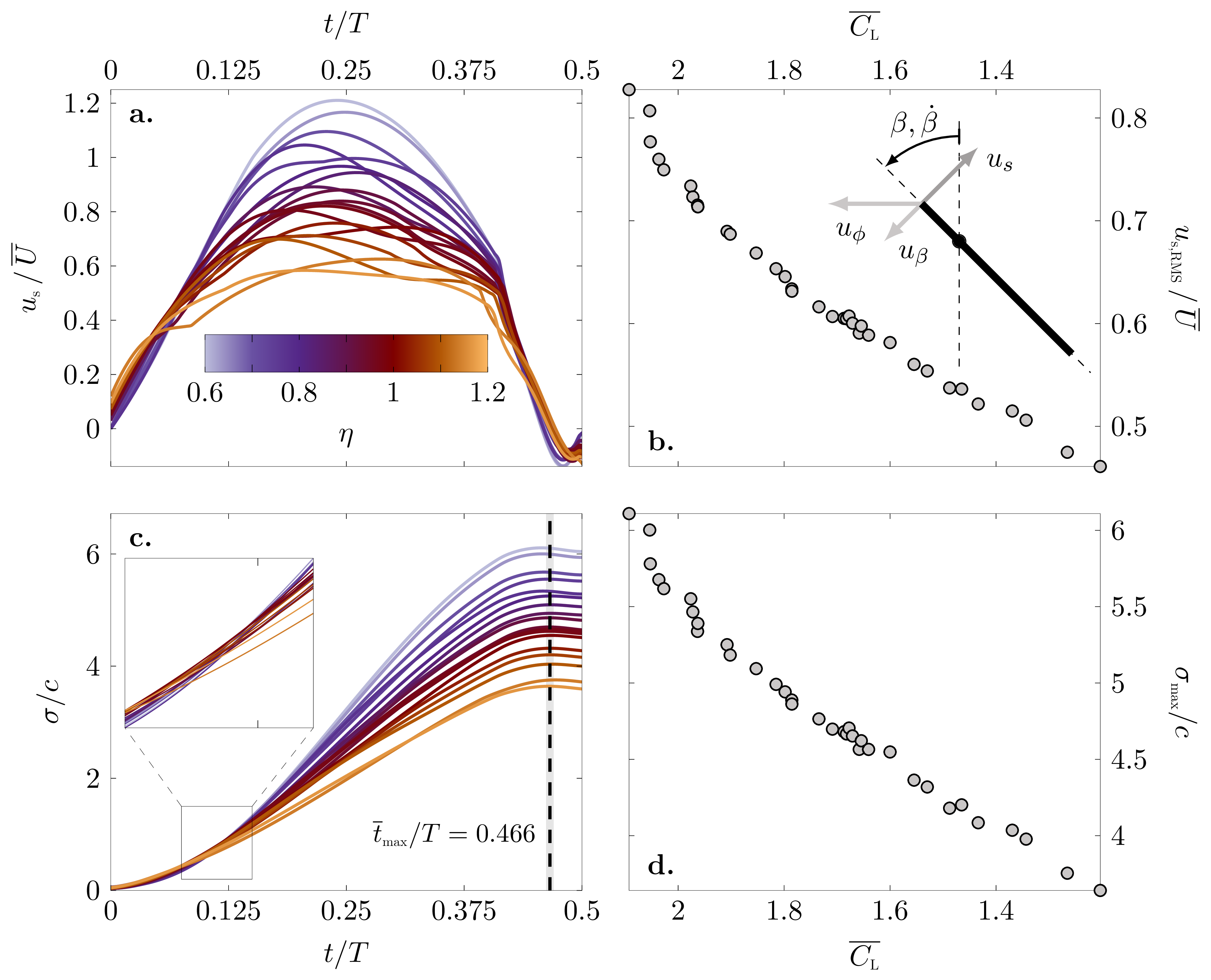}
		\caption{a.\@ Instantaneous shear layer velocity \kindex{u}{s} over time $t/T$, b.\@ RMS of the shear layer velocity \kindex{u}{s,RMS} over $\kindex{\overline{C}}{L}$, and c.\@ chord-normalized advective time $\sigma / c$ over time $t/T$, and d.\@ advective time maxima $\kindex{\sigma}{max}$ over $\kindex{\overline{C}}{L}$.
			Color-coded is the hovering efficiency $\eta$ corresponding to the Pareto front individual.}
		\label{fig:shear_layer_velo}
	\end{figure}

\begin{figure}[tb]
		\centering
		\includegraphics[]{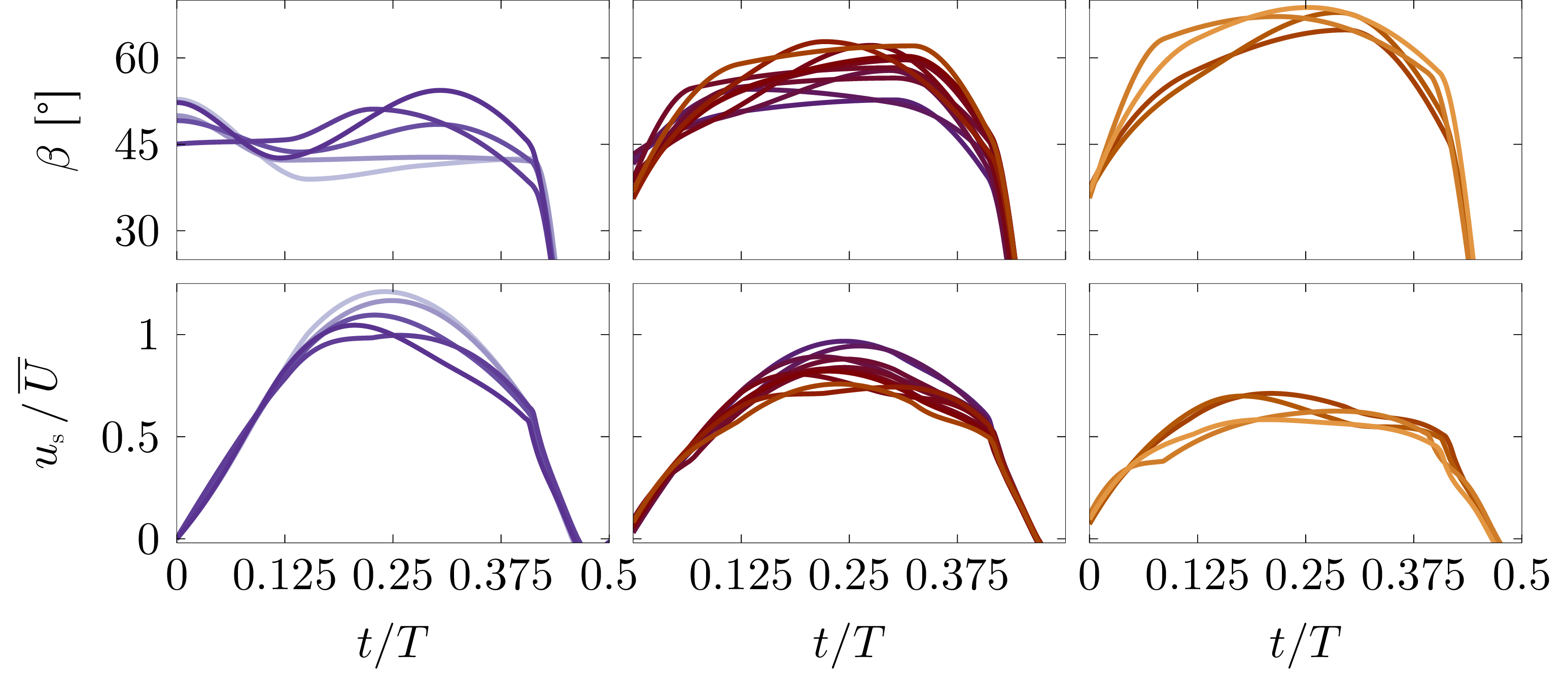}
		\caption{Leading edge shear layer velocity \kindex{u}{s} profile groups derived from pitch angle $\beta$ kinematic groups}
		\label{fig:us_types}
	\end{figure}

	\Cref{fig:shear_layer_velo}a summarises the temporal evolution of the leading edge shear layer velocity \kindex{u}{s} for all pitch angle kinematics along the Pareto front in \cref{fig:pareto_front}.
	The leading edge shear layer velocity is non-dimensionalised by the average stroke velocity $\overline{U}$.
	The shear layer velocity profiles can again be divided into three characteristic groups based on their form (\cref{fig:us_types}).
	The three groups correspond to the same main middle portion of the Pareto front and the high lift and high efficiency tails.
	The solutions with trapezoidal pitch angle profiles that yield maximum \kindex{\overline{C}}{L} have sinusoidal \kindex{u}{s}-evolutions.
	The pitch angle $\beta$ is approximately constant during large portions of the stroke cycle and the shear layer velocity is mainly driven by the stroke velocity.
	The most efficient solutions with sinusoidal pitch angle profiles have more trapezoidal shear layer velocity profiles.
	Here, the pitch angle decreases when the stroke velocity increases and vice versa to obtain an approximately constant value of the shear layer velocity during most of the stroke.
	The shear layer velocity profiles for the intermediate solutions gradually evolve from the sinusoidal shape with high maximum values around mid stroke to the trapezoidal shapes with a rounded ascending flank and a plateau at lower values.

	To scale the aerodynamic performance of the flapping wing hovering motion it is desirable to have a single characteristic velocity which is representative of the input kinematics.
	We propose to use the root-mean-square value of the shear layer velocity \kindex{u}{s,rms} which serves as a fundamental measure of the magnitude of an oscillating signal.
	The root-mean-square value of the shear layer velocity decreases continuously along the entire Pareto front with decreasing stroke average lift~\cref{fig:shear_layer_velo}b.
	This single kinematic parameter allows for the sorting of the aerodynamic performance of the kinematics in terms of the two objectives of the optimisation: mean lift and efficiency.

	The temporal evolution of the advective time $\sigma$ for all pitch angle kinematics along the Pareto front is summarised in \cref{fig:shear_layer_velo}c.
	The advective time has the dimension of length and is non-dimensionalised by the chord length.
	According to \cref{eq:adv_time1}, the advective time is zero at the start of the pitching cycle and increases monotonically until the shear layer velocity becomes negative following the initiation of the pitch rotation near the end of the stroke ($t/T\approx0.42$).
	The shape of the advective time curves is similar for all solutions.
	The advective time evolutions of solutions that yield higher mean lift are above those that are more efficient.
	This is true at any time beyond $t/T=0.125$.
	Prior to $t/T=0.125$, the more efficient kinematics have a higher shear layer velocity due to a faster pitch rotation and higher advective times (\cref{fig:shear_layer_velo}c).
	The more efficient kinematics typically have lower angles of attack during most of the stroke motions and require a more important pitch rotation around stroke reversal.
	The sign-reversal of the shear layer velocity at the end of the stroke motion marks the end of the feeding cycle of the current leading edge vortex.
	The maximum advective time indicates the age of the leading edge vortex at the end of the feeding cycle.
	The leading edge vortex created by the highest lift generating kinematics reaches a vortex age of \num{6.11} advective time scales before the pitch rotation sets in.
	The maximum advective time $\kindex{\sigma}{max}$ decreases with decreasing mean lift $\kindex{C}{L}$ along the entire Pareto front (\cref{fig:shear_layer_velo}d).
	The leading edge vortex created by the most efficient kinematics only reaches a vortex age of \num{3.64} advective time scales.
	\Cref{fig:shear_layer_velo}d reveals a direct relationship between the maximum age of the leading edge vortex and the aerodynamic performance of the hovering motion.
	This inspires us to use to the advective time as the characteristic time scale to scale the temporal evolution of the aerodynamic response to the various flapping wing hovering kinematics.

	In the following, we will demonstrate how the root-mean-square value of the shear layer velocity and the advective time can be respectively used as characteristic velocity and time to scale the temporal evolution of the leading edge vortex circulation and the aerodynamic performance of Pareto-optimal the flapping wing kinematics.

	\subsubsection{Leading edge vortex circulation}
\begin{figure}[tb]
		\centering
		\includegraphics[width=\textwidth]{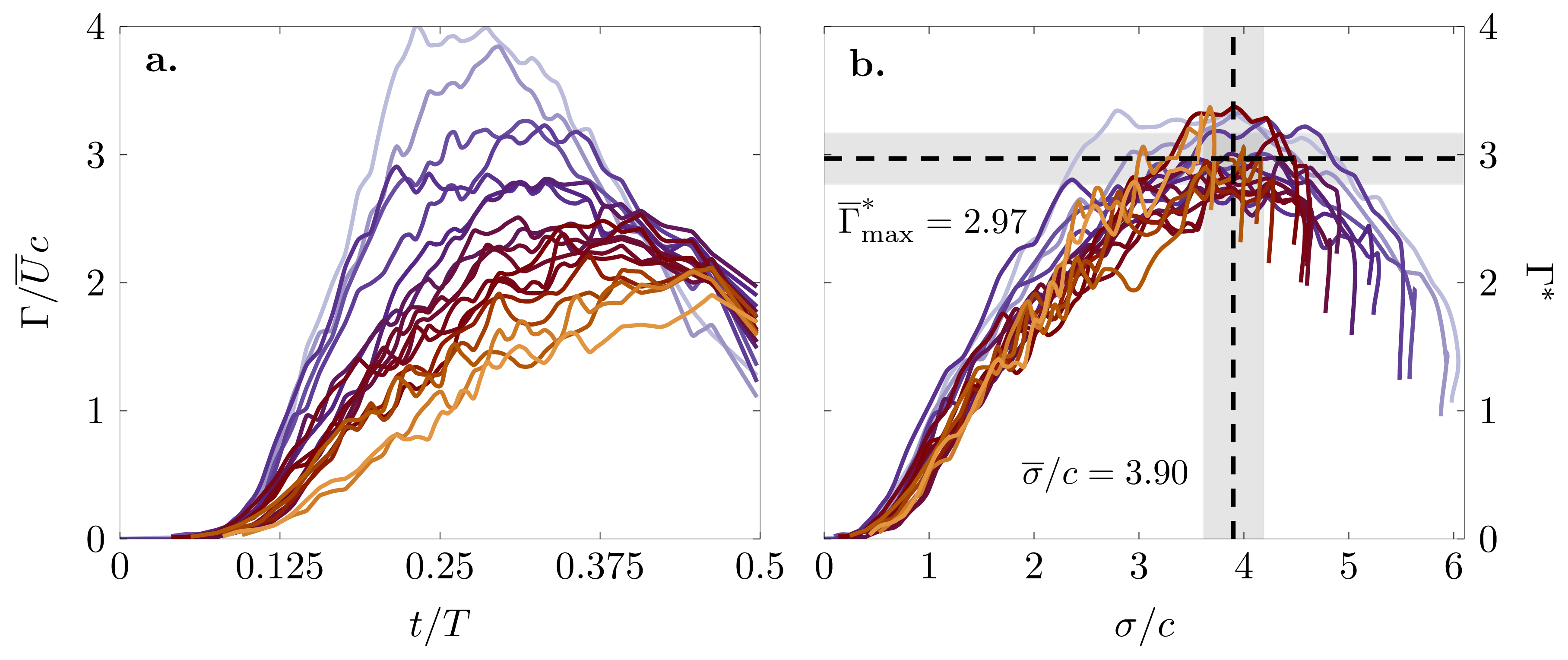}
		\caption{a\@. Normalized leading edge vortex circulation $\Gamma / \overline{U} c$ over time $t/T$, b\@. leading edge vortex circulation $\Gamma / \overline{u}_s c$ scaled with $\hat{u}_s$ over advective time $\sigma / c$.
			Color-coded is the hovering efficiency $\eta$ corresponding to the Pareto front individual.
			The dashed lines mark the mean of the scaled circulation $\Gamma^*$ maxima and the corresponding mean timing $\overline{\sigma}/c$.
			The gray areas represents +/- one standard deviation around the mean.}
		\label{fig:circulation}
	\end{figure}

	\Cref{fig:circulation} shows a comparison of temporal evolution of the leading edge vortex circulation $\Gamma$ for all solutions along the Pareto front for two different normalisations.
	In \cref{fig:circulation}a, the circulation is normalised by the stroke average velocity $\overline{U}$ and the chord length and the time axis is normalised by the flapping period.
	Note that the stroke average velocity and the flapping period are the same for all kinematics considered here.
	In \cref{fig:circulation}b, the circulation is normalised by maximum leading edge shear layer velocity \kindex{u}{s,max} and the chord length and is presented as a function of the non-dimensionalised advective time $\sigma/c$.

	The leading edge vortex circulation curves all start at zero and start to increases when a new vortex starts to emerge near the wing's leading edge.
	The circulation increases as the leading edge vortex grows and reaches a maximum value at some point during the second part of the stroke cycle depending on the pitch angle kinematics.
	Kinematics that yield higher $\kindex{\overline{C}}{L}$, generate circulation at a higher rate, reach a higher maximum value of the circulation earlier in the flapping cycle (\cref{fig:circulation}a).
	The peak circulation for the highest $\kindex{\overline{C}}{L}$ motion is reached around mid-stroke  when the maximum stroke velocity is reached.
	The circulation for the most efficient motion continues to increase until the pitch rotation sets in at the end of the stroke motion.
	The peak value gradually decreases and the timing of the peak gradually delays when we move along the Pareto front sacrificing lift for efficiency.

	If the circulation is now presented as a function of the advective time as defined in \cref{eq:adv_time1} and normalised by the maximum leading edge shear layer velocity and the chord length, all curves collapse and follow the same trajectory (\cref{fig:circulation}b).
	The newly scaled circulation $\Gamma^* = \Gamma/(\kindex{u}{s,max}\,c)$ reaches a maximum value of $\kindex{\Gamma^*}{max}\approx 3$ after $\overline{\sigma}/c = 3.90$.
	For all pitching kinematics along the Pareto front, the maximum leading edge vortex circulation scales with the maximum local shear layer velocity and this maximum circulation is reached after the leading edge has traveled a distance of four chord lengths regardless of the temporal evolution of the pitch angle during the flapping motion.

	The optimal kinematics are tailored to reach the maximum circulation right before starting the pitch rotation near stroke reversal.
	The high lift kinematics continue after the maximum leading edge vortex circulation is reached and cover more advective times during a stroke cycle.
	For $\sigma/c>4$, the vortex circulation decreases even though the vortex stays close to the wing.
	During this part of the motion, vorticity continues to be produced and fed through the leading edge shear layer without increasing the leading edge vortex circulation.
	This vorticity must be transported either in span-wise direction or dissipates as a consequence of vortex bursting.

	This scaling of the leading edge vortex circulation based on the maximum shear layer velocity was previously demonstrated to be effective for two-dimensional starting vortices~\cite{sattari_growth_2012} and swept and unswept pitching wings~\cite{Onoue_Breuer_2017}.
	The constant vortex formation time of approximately four advective times is also consistent with many examples of optimal vortex formation found in nature~\cite{Dabiri_2009} and with
	the many records of vortex formation numbers around four for vortex rings generated by a piston cylinders~\cite{gharib_universal_1998}.

	%
	%
	%
	\subsubsection{Aerodynamic loads}
	%
	The leading edge vortex provides an important contribution to the aerodynamic forces on unsteadily moving wings~\cite{Eldredge_Jones_2018}.
	The evolution of the leading edge vortex circulation $\Gamma$ in the measurement plane at $R_2$ scales in magnitude with the maximum leading edge shear layer velocity \kindex{u}{s,max}.
	Based on this new scaling of the circulation and the Kutta-Joukowski theorem, we can formulate the sectional lift $L$ as:
	\begin{equation}
	L=\rho \overline{U} \Gamma = \rho \overline{U} \Gamma^* \kindex{u}{s,max} c \quad,
	\end{equation}
	and a rescaled lift coefficient $\kindex{C}{L}^*$ as:
	\begin{equation}
	\kindex{C}{L}^*=\frac{L}{1/2 \rho \overline{U} \kindex{u}{s,rms} Rc}\quad.
	\label{eq:CL_us}
	\end{equation}
	Here, we have replaced the maximum shear layer velocity \kindex{u}{s,max} by the root-mean-square value of the shear layer velocity \kindex{u}{s,rms} to better account for the span-wise variation of the shear layer velocity.

\begin{figure}[tb]
		\centering
		\includegraphics[]{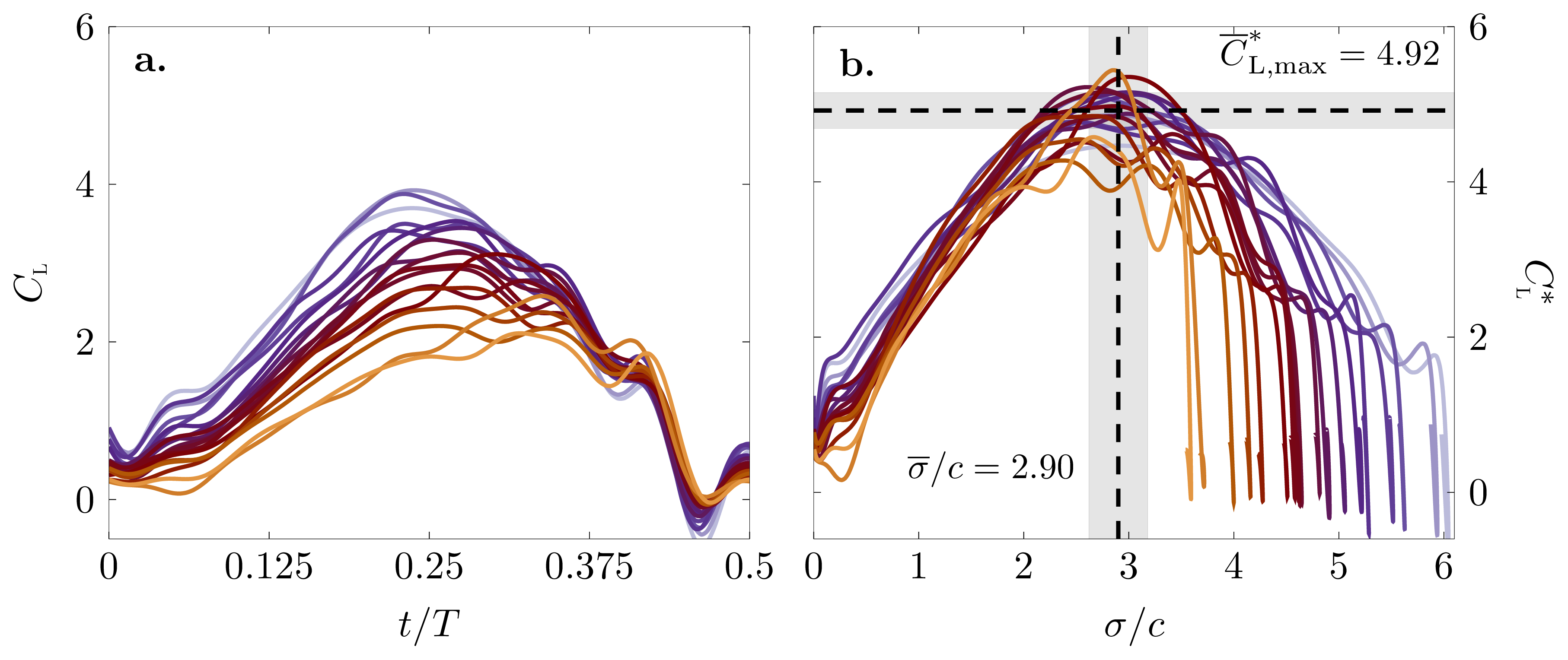}
		\caption{a.\@ Lift coefficient $\kindex{C}{L}$ over time $t/T$ and b.\@ scaled lift coefficient $C^*_{L}$ over advective time $\sigma / c$.
			Color-coded is the hovering efficiency $\eta$ corresponding to the Pareto front individual.
			The dashed lines mark the mean of the scaled lift $C^*_{L}$ maxima and the corresponding mean timing $\overline{\sigma}/c$.
			The gray areas represents +/- one standard deviation around the mean.}
		\label{fig:CL_scaling}
	\end{figure}

	The comparison of this new scaling of the lift coefficient in comparison to the more standardly used definition $\kindex{C}{L}=L/\left(1/2 \rho \overline{U}^2 Rc \right)$ is presented in \cref{fig:CL_scaling} for all solutions along the Pareto front.
	The maximum values of lift coefficient \kindex{C}{L} in \cref{fig:CL_scaling}a decrease and occur later in the cycle for kinematics that are more efficient.
	When the lift coefficient is normalised according to \cref{eq:CL_us} and presented as a function of the non-dimensionalised advective time $\sigma/c$ in \cref{fig:CL_scaling}b the increasing lift slopes collapse and the magnitude and timing of the lift coefficient maxima align.
	With the proposed scaling, the lift coefficient reaches a maximum value around $\kindex{C}{L,max}^* = 4.92$ for $\sigma/c = 2.90$ for all Pareto-optimal kinematics.
	The lift coefficient maximum is reached one advective time before the leading edge vortex circulation $\Gamma$ reaches its maximum value.
	This indicates that $\kindex{C}{L,max}^*$ depends not only on the strength of the leading edge vortex, but also on its position with respect to the wing.
	The timing of both scales with the advective time for all kinematics considered here.

	\subsubsection{Hovering efficiency}
	Analogous to the leading edge vortex circulation $\Gamma$ and the lift coefficient \kindex{C}{L}, the drag \kindex{C}{D} and power coefficient \kindex{C}{P} can be renormalised using the leading edge shear layer velocity to:
	\begin{equation}
	\kindex{C}{D}^* = \frac{D}{\frac{1}{2} \rho \overline{U} \kindex{u}{s,rms} R c}
	\quad \textnormal{and} \quad
	\kindex{C}{P}^* = \frac{P}{\frac{1}{2} \rho \overline{U} \kindex{u}{s,rms}^2 R c}
	\label{eq:CD_CP_us}
	\end{equation}
	Using the coefficients $\kindex{C}{L}^*$ and $\kindex{C}{P}^*$, we rescale the stroke average hovering efficiency $\eta$ :
	\begin{equation}
	\eta^* = \frac{\kindex{\overline{C}}{L}^*}{\kindex{\overline{C}}{P}^*} = \frac{\overline{L}}{\overline{P}} \, \kindex{u}{s,rms}\quad.
	\label{eq:eta_us}
	\end{equation}

\begin{figure}[tb]
		\centering
		\includegraphics[]{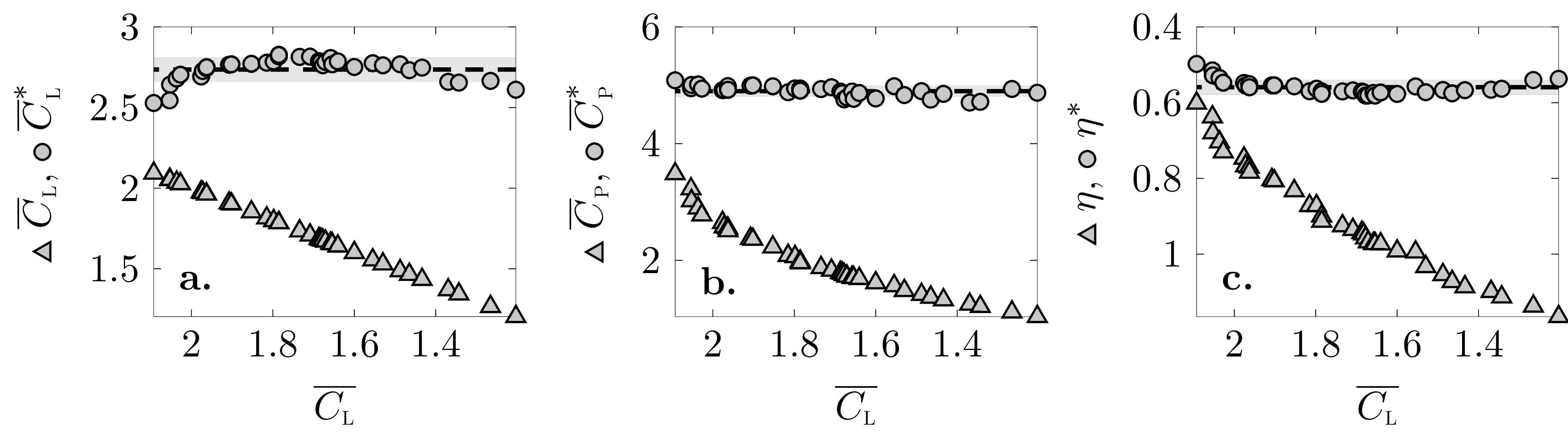}
		\caption{a.\@ Scaled vs. unscaled stroke-average lift coefficient \kindex{\overline{C}}{L} over \kindex{\overline{C}}{L},
			b.\@ scaled vs. unscaled stroke-average power coefficient \kindex{\overline{C}}{P} over \kindex{\overline{C}}{L},
			and c.\@ scaled vs. unscaled stroke-average hovering efficiency $\eta$ over \kindex{\overline{C}}{L}.
			The triangles represent the normalized aerodynamic coefficients and the circles are the coefficients rescaled with the shear layer velocity \kindex{u}{s,rms}.
			The dashed lines mark the mean of the scaled values.
			The grey areas represents +/- one standard deviation around the mean.}
		\label{fig:eta_scaling}
	\end{figure}

	\Cref{fig:eta_scaling} shows the comparison of the newly scaled stroke average lift, power, and the efficiency with the standard normalisation for all solutions along the Pareto front.
	The stroke average lift and power coefficients \kindex{\overline{C}}{L} and \kindex{\overline{C}}{p} normalised based solely on the stroke average velocity $\overline{U}$ in \cref{fig:eta_scaling}a,b increase with increasing \kindex{\overline{C}}{L}.
	Note that we have kept the y-axis inverted here to match the Pareto front representation in \cref{fig:eta_scaling}c\@.
	When we normalise the coefficients as proposed in \cref{eq:CL_us,eq:CD_CP_us} using the root-mean-square value of the shear layer velocity, the scaled coefficients collapse and we obtain mean values of $\kindex{\overline{C}}{L}^* = 2.74$ and $\kindex{\overline{C}}{P}^* = 4.90$ across the ensemble of Pareto-optimal kinematics.
	The rescaled hovering efficiency $\eta^*$ reaches an average value of $\overline{\eta}^* = 0.56$ for all Pareto front solutions.
	The standard deviation around these mean values across all kinematics is indicated by the grey shading in \cref{fig:eta_scaling}.

	The proposed scaling works especially well for the power coefficient.
	Small deviations from the constant mean values of the lift coefficient and efficiency are observed for the most efficient kinematics and the kinematics that yield the highest lift.
	The successful scaling of the efficiency with the shear layer velocity confirms the strong correlation between the aerodynamic efficiency and the growth rate of the leading edge vortex for the Reynolds number considered in this work.

	Based on this scaling, the shape of the Pareto-front would change from a convex $\eta$ versus $\kindex{\overline{C}}{L}$ shape, to basically a single point in the $\overline{\eta}^*$ versus $\kindex{\overline{C}}{L}^*$ plane.
	The values corresponding to this optimal performance point are expected to vary for different wing planforms, different flow conditions, Reynolds number, and reduced frequency.
	The solutions that end up in the optimal performance point are not unique, they all create a leading edge vortex of a different size and strength, but do this in the most optimal way.
	This optimal vortex formation process is governed by the shear layer velocity which serves as the main indicator of the aerodynamic performance.
	The scaled values of non-Pareto optimal kinematics do not collapse and are lower than the $\kindex{\overline{C}}{L}^*$ and $\overline{\eta}^*$ values, and higher than the $\kindex{\overline{C}}{P}^*$ values found in \cref{fig:eta_scaling}.
	Higher values $\kindex{\overline{C}}{L}^*$ and $\overline{\eta}^*$ and lower $\kindex{\overline{C}}{P}^*$ values are not achievable with the given geometric and kinematic boundary conditions and the current values obtained by the scaled coefficients represent target limits for optimal performance.
	The limiting values can be used to quickly estimate the maximal achievable performance values of new or adapted kinematics using
	\begin{equation}
	\kindex{\overline{C}}{L}(\phi,\beta) = \kindex{\overline{C}}{L}^* \; \frac{\kindex{u}{s,rms}}{\overline{U}} \quad ,
	\label{eq:CL_prediction}
	\end{equation}
	\begin{equation}
	\kindex{\overline{C}}{P}(\phi,\beta) = \kindex{\overline{C}}{P}^* \; \frac{\kindex{u}{s,rms}^2}{\overline{U}^2} \quad ,
	\label{eq:CP_prediction}
	\end{equation}
	\begin{equation}
	\eta(\phi,\beta) = \overline{\eta}^* \; \frac{\overline{U}}{\kindex{u}{s,rms}} \quad ,
	\label{eq:eta_prediction}
	\end{equation}
	with $\kindex{\overline{C}}{L}^* = 2.74$, $\kindex{\overline{C}}{P}^* = 4.90$ and $\overline{\eta}^* = 0.56$ respectively.
This estimation is possible without additional measurements because the shear layer velocity only depends on the input kinematics ($\phi$, $\beta$).
	Further investigations are desirable to determine how the values of $\kindex{\overline{C}}{L}^*$, $\kindex{\overline{C}}{P}^*$, and $\overline{\eta}^*$ vary as function of the Reynolds number and wing geometry.

	\section{Conclusion}%
	We have experimentally optimised the pitch angle kinematics for hovering flapping wing flight using a unique mechanical flapping wing system that allows for robust and repeatable execution of widely varying flapping kinematics.
The kinematics yielding maximal stroke average lift and hovering efficiency have been determined with the help of an evolutionary algorithm and in-situ force and torque measurements at the wing root.
	Additional flow field measurements have been conducted to reveal the phenomenology of the force and flow field response for the Pareto-optimal kinematics.

	A Pareto front of optimal solutions is obtained along which the stroke-average lift $\kindex{\overline{C}}{L}$ produced by the optimised kinematics ranges from \numrange{1.20}{2.09} and the aerodynamic performance $\eta$ varies from \numrange{0.60}{1.17}.
	By trading off up to \SI{43}{\percent} of its maximum lift capacity, the flapping wing system's efficiency can be increased by \SI{93}{\percent} by merely adjusting the pitch angle kinematics.

	The Pareto-optimal pitching kinematics are classified into three groups with distinctly different kinematic and dynamic characteristics.
	The three groups correspond to three sections of the Pareto front: the large central part where the lift increases approximately linearly with decreasing efficiency, and the high lift and high efficiency tails where there is a significantly larger trade off between lift and efficiency.
	The kinematics in the high-lift tail of the Pareto front have a trapezoidal pitch angle profile with a plateau around $\beta=\ang{45}$ and an additional peak at the beginning of the stroke.
	These kinematics create strong leading edge vortices early in the cycle which enhance the force production on the wing.
	The leading edge vortex circulation and the aerodynamic forces reach their maxima around mid-stroke marking the the end of the growth of the leading edge vortex and the onset of vortex lift-off.
	The transition between the different phases of the leading edge vortex development are identified based on the surface velocity, the trajectory of the surface half saddles from FTLE ridges, and the position of the leading edge vortex with respect to the wing.
	The most efficient kinematics have a more rounded, sinusoidal profile with a maximum pitch angle $\beta>\ang{60}$ around mid-stroke and create weaker leading edge vortices that stay close-bound to the wing throughout the majority of the stroke-cycle.
	The aerodynamic forces and the leading edge vortex circulation grow significantly slower in the high efficiency tail than in the rest of the Pareto front and reach their maxima just before the end-of-stroke rotation.
	The efficient leading edge vortex development is characterised by the absence of vortex lift-off.
	The kinematics in the bulk of the Pareto front gradually evolve from the more trapezoidal high lift kinematics toward the almost sinusoidal high efficient kinematics.
	The aerodynamic forces and leading edge vortex circulation reach maximum values shortly after mid-stroke.

The classification into three groups also applied to the evolution of the shear layer velocity which is directly determined from the input kinematics.
Kinematics within the same group yield similar characteristic evolutions of the shear layer velocity that are different from those of the other groups.
	The integral of time-resolved leading edge shear layer velocity \kindex{u}{s} over the cycle time $t$ yields the advective time $\sigma$ which serves as a normalised time scale for the leading edge vortex growth and aerodynamic force evolution.
The root-mean-square value of the shear layer velocity at the leading edge serves to quantitatively characterise the growth of the leading edge vortex and scale the average and the temporal evolutions of the circulation and the aerodynamic forces.
	The ascending flanks and maxima of the leading edge circulation $\Gamma$ and lift coefficient \kindex{C}{L} collapse when being normalised by the root-mean-square value of the shear layer velocity and presented in function of the advective time for all Pareto front kinematics.
	The optimal kinematics are tailored to reach the maximum circulation right before starting the end-of-stroke pitch rotation after $\overline{\sigma}/c = 3.9$.
	The high lift kinematics continue after	the maximum leading edge vortex circulation is reached and cover more advective times during a stroke cycle.
	The vortex formation time of approximately four advective times for the most efficient hovering kinematics is consistent with many examples of optimal vortex formation found in nature.

	The leading edge shear layer velocity \kindex{u}{s,RMS} also serves to renormalise the aerodynamic power coefficient \kindex{C}{P} and hovering efficiency $\eta$.
	All cycle-average aerodynamic coefficients normalised by \kindex{u}{s,RMS} collapse onto their mean-values $\kindex{\overline{C}}{L}^* = 2.74$, $\kindex{\overline{C}}{P}^* = 4.90$ and $\overline{\eta}^* = 0.56$ for every Pareto front kinematic.
	The	successful newly proposed scaling of the efficiency with the shear layer velocity confirms the strong correlation between the aerodynamic efficiency and the growth rate of the leading edge vortex for the Reynolds number considered in this work.
The correlation is based on the underlying physics and we expect the general phenomenology and the scaling based on the shear layer velocity to be valid for different wing shapes and even flexible wings.
	Furthermore, the shear layer velocity is determined solely on the basis of the input kinematics and this scaling allows us to estimate the maximally attainable stroke-average lift, power, and efficiency of new or adapted kinematics.

	Further investigations are desirable to	determine how the values of $\kindex{\overline{C}}{L}^*$, $\kindex{\overline{C}}{P}^*$ and $\overline{\eta}^*$ vary as function of the Reynolds number, different wing planforms and for non Pareto-optimal kinematics.
	Taking into account the larger variations of kinematics considered here and the three-dimensionality of the flapping wing motion, the robustness of the proposed scaling is remarkable and can guide the aerodynamic design of human-engineered devices that can automatically adapt their motion kinematics to optimally fit varying flight conditions.
The results should also be transferable to other unsteady aerodynamic problems that are vortex dominated and where the vortex is accumulating circulation resulting from an arbitrary relative unsteady motion of an aerodynamic body.

	\section{Author Contributions}%
	A.G. implemented the genetic algorithm-based optimisation, designed, and conducted the experiments.
K.M. acquired the funding for the project, conceptualised and supervised the project.
	Both authors contributed to the writing of the manuscript and to the analysis of the results.
	\section{Acknowledgments}%
	This work was supported by the Swiss National Science Foundation under grant number 200021\_175792
	\section{References}%
	\bibliography{ms_v2}
	\bibliographystyle{ieeetr}
\end{document}